\documentclass[aps,eqsecnum,superscriptaddress,nofootinbib,preprint]{revtex4}

\usepackage{amsmath} 
\usepackage{amssymb} 
\usepackage{dsfont} 
\usepackage{mathrsfs} 
\usepackage{eufrak} 
\usepackage{wasysym} 
\usepackage{tikz-cd} 
\usepackage{mathtools} 

\usepackage{boondox-cal}

\usepackage{footnotebackref} 

\newcommand{\iprod}{\mathbin{\lrcorner}} 

\newcommand{\secref}[1]{Sec.~\ref{#1}}
\newcommand{\figref}[1]{Fig.~\ref{#1}}
\newcommand{\footref}[1]{Footnote~\ref{#1}}

\begin{document}

\title{A unified view of curvature and torsion in metric-affine gauge theory of gravity through affine-vector bundles}

\author{Bo-Hung Chen}
\email{kenny81778189@gmail.com}
\affiliation{Department of Physics, National Taiwan University, Taipei 10617, Taiwan}
\affiliation{Center for Theoretical Physics, National Taiwan University, Taipei 10617, Taiwan}

\author{Dah-Wei Chiou}
\email{dwchiou@gmail.com}
\affiliation{Department of Physics, National Sun Yat-sen University, Kaohsiung 80424, Taiwan}
\affiliation{Center for Condensed Matter Sciences, National Taiwan University, Taipei 10617, Taiwan}


\begin{abstract}
One of the most appealing results of metric-affine gauge theory of gravity is a close parallel between the Riemann curvature two-form and the Cartan torsion two-form: While the former is the field strength of the Lorentz-group connection one-form, the latter can be understood as the field strength of the coframe one-form. This parallel, unfortunately, is not fully established until one adopts Trautman's idea of introducing an affine-vector-valued zero-from, the meaning of which has not been satisfactorily clarified. This paper aims to derive this parallel from first principles without any ad hoc prescriptions. We propose a new mathematical framework of an \emph{associated affine-vector bundle} as a more suitable arena for the affine group than a conventional vector bundle, and rigorously derive the covariant derivative of a local section on the affine-vector bundle in the formal Ehresmann-connection approach. The parallel between the Riemann curvature and the Cartan torsion arises naturally on the affine-vector bundle, and their geometric and physical meanings become transparent. The clear picture also leads to a conjecture about a kinematical effect of the Cartan torsion that in principle can be measured \textit{\`{a} la} the Aharonov-Bohm effect.
\end{abstract}


\maketitle

\section{Introduction}\label{sec:introduction}
Gauge theories of gravity are the effort to cast gravitation in the language of Yang-Mills theory. Various approaches of gauge theories of gravity, differing from one another by considering different local gauge groups, have been intensively developed in the past 40 years, and they have revealed various profound geometric structures of spacetime and achieved many inspiring results. (See \cite{blagojevic2002gravitation,mielke2017geometrodynamics} for comprehensive accounts; also see \cite{gronwald1996gauge} for a brief review and \cite{blagojevic2013gauge} for a historical account.)
Even though a satisfactory gauge theory of gravity remains elusive, undertaking of the research has enormously enhanced our knowledge about many fundamental issues --- inertial effects of a spin particle \cite{hehl1990inertial,obukhov2013spin}, relationship between spin and torsion \cite{hehl1976general}, nonlinear effects of gravity, topological aspects of gravity, to name a few \cite{blagojevic2002gravitation, mielke2017geometrodynamics, gronwald1996gauge}.

Among various gauge approaches of gravity, metric-affine gauge theory (MAG) of gravity is a well-developed formulation that has many intriguing features (see \cite{gronwald1996gauge,hehl1995metric} for reviews). In MAG, the gauge group is given by the affine group $A(n,\mathbb{R}):=\mathbb{R}^n\rtimes GL(n,\mathbb{R})$ --- i.e.\ the semidirect product of the vector space $\mathbb{R}^n$ and general linear group $GL(n,\mathbb{R})$, and correspondingly the $\mathfrak{a}(n,\mathbb{R})$-valued affine connection $\mathcal{A} = \Gamma^{(L)} + \Gamma^{(T)}$ plays the role of a gauge potential, where $\Gamma^{(T)}$ is associated with the \emph{translational} subgroup $T(n,\mathbb{R})\cong\mathbb{R}^n$ of the affine group and $\Gamma^{(L)}$ with the \emph{linear} subgroup $GL(n,\mathbb{R})$.
Particularly, if the general linear group $GL(n,\mathbb{R})$ is replaced by the Lorentz group $SO(1,n-1)$, MAG is reduced to the Poincar\'{e} gauge theory of gravity.

The result of MAG profoundly suggests that the Cartan torsion two-form $T$ and the Riemann curvature two-form $R$ can be understood on an equal footing. More precisely, we have
\begin{subequations}\label{T and R}
\begin{eqnarray}
\label{T and R part a}
T &=& D^{(L)}\theta \equiv d\theta + \Gamma^{(L)}\wedge \theta,\\
\label{T and R part b}
R &=& D^{(L)}\Gamma^{(L)} \equiv d\Gamma^{(L)} + \Gamma^{(L)}\wedge\Gamma^{(L)},
\end{eqnarray}
\end{subequations}
where $\theta$ is the coframe one-form, $\Gamma^{(L)}$ is the gauge potential one-form of the group $GL(n,\mathbb{R})$, and $D^{(L)}$ is the exterior covariant derivative associated with $\Gamma^{(L)}$. These draw a close parallel between $T$ and $R$ in the sense that the former is the field strength of $\theta$ while the latter the field strength of $\Gamma^{(L)}$.\footnote{This parallel can also be extended to the \emph{nonmetricity} field $Q$, which can be viewed as the field strength of the metric tensor. See Table~5 in \cite{gronwald1996gauge} and Table~1 in \cite{hehl1995metric}. For our purpose, we focus on the parallel between $T$ and $R$ only.} It is tantalizing to equate $\theta$ with $\Gamma^{(T)}$ and furthermore to identify $T$ and $R$ as the translational part $R^{(T)}$ and the linear part $R^{(L)}$, respectively, of the affine gauge curvature as defined in \eqref{cal R}. Unfortunately, while $R$ is to be identified as $R^{(L)}$, the fact that $\theta$ and $\Gamma^{(L)}$ transform differently under the $A(n,\mathbb{R})$ gauge transformation spoils the beauty of identifying $\theta$ as $\Gamma^{(L)}$ and $T$ as $R^{(T)}$.
To resolve this glitch, Trautman \cite{trautman1973structure} proposed a solution that introduces an \emph{affine-vector}-valued zero-form $\xi^i$ and defines the new one-form as
\begin{equation}\label{theta and xi}
  \theta:=\Gamma^{(T)}+D^{(L)}\xi.
\end{equation}
The new one-form $\theta$ then can be identified as the coframe field, and correspondingly $T$ and $R^{(T)}$ are related via
\begin{equation}\label{T and RT}
  T^i=R^{(T)i}+R_j^{(L)i}\xi^j.
\end{equation}

Tremendous effort has been devoted to interpreting $\xi$ and deriving it from different perspectives.\footnote{Meanwhile, there have been some approaches that apparently do not get involved with $\xi$ but nevertheless give rise to the coframe one-form $\theta$. For example, in \cite{sharpe2000differential,sternberg2013curvature}, the approach of a ``reductive principle bundle'' was proposed, whereby $\Gamma^{(T)}$ is directly identified as $\theta$ without referring to $\xi$. However, this approach does not seem to manifest the complete symmetry of the affine group $A(n,\mathbb{R})$, obscuring the close parallel
between the Riemann curvature and the Cartan torsion.}
The field $\xi$ also appears in the context of gauged nonlinear realizations of the translation group \cite{kawai1986overline, kawai1991extended, lord1987unified, lopez1995ordinary, julve1996nonlinear}.
By requiring $\Gamma^{(T)}=0$, $\xi$ can be interpreted as Cartan's ``generalized radius vector'' \cite{cartan1986manifolds,mielke1993avoiding}.
By requiring $D^{(L)}_\mu\xi^i=\delta^i_\mu$, which is used to connect the coset space $A(n,\mathbb{R})/GL(n,\mathbb{R})\cong\mathbb{R}^n$ to the cotangent space of the spacetime manifold, $\xi$ can be kinematically understood as the ``Poincar\'{e} coordinates'' \cite{hayashi1967extended, hayashi1980gravity, shirafuji1988gauge, muller1984gauge, ivanov1980gauge, grignani1992gravity}.
The field $\xi$ could also be explained in view of the theory of dislocations \cite{sardanashvily1987dislocation} or in terms of jet bundles \cite{hennig1981gravity}.
The fact that $\Gamma^{(T)}$ is not directly identical to $\theta$ might give rise to a gravitationally induced geometric phase \cite{morales1995geometrical}.
Despite unceasing endeavors to understand $\xi$ and its relation to translational symmetry, ``the story of $\xi$ has not yet come to an end \cite{hehl1995metric}'', and ``the geometric and physical meaning of the relation [\eqref{theta and xi}], especially the role of the field $\xi$, is perhaps not completely satisfactorily clarified, yet \cite{gronwald1996gauge}.''

Furthermore, it should be noted that the parallel between $T$ and $R$ has not been completely spelled out. It is well known that the Riemann curvature yields two geometric consequences: \emph{geodesic deviation} and \emph{holonomy around a closed curve}. Correspondingly, the Cartan torsion is expected to have the two analogous consequences as well.
Geodesic deviation describes the tendency of bending towards or away from each other of two neighboring geodesics that are initially parallel to each other (see Ch.~11 of \cite{Misner:1974qy} for a detailed account).
Analogously, the Cartan torsion indicates the tendency of \emph{closure failure} of a parallelogram --- i.e., it gives rise to a displacement vector, which describes how much the two endpoints of an infinitesimal parallelogram fail to coincide (see \cite{gronwald1996gauge}, especially Fig.~4 therein, for a detailed account).
The Riemann curvature and the Cartan torsion are analogous to each other in the sense that both of them measure how much the curved spacetime is deviated from the flat one in regard to parallelogram distortion, albeit from different considerations.

On the other hand, $R(X,Y)$ yields a value of the general linear algebra $\mathfrak{gl}(n,\mathbb{R})$, which corresponds to the holonomy around an infinitesimal closed curve spanned by the two vectors $X$ and $Y$ as depicted in \figref{fig:holonomy loop}. The holonomy of $R$ is well understood as a linear transformation that tells the difference between the initial and final states for a given vector that is parallel transported around the closed curve until it is back to the starting point (see Ch.~11 of \cite{Misner:1974qy} for a detailed account).
Likewise, $T(X,Y)$ yields a value of the translation algebra $\mathfrak{t}(n,\mathbb{R})$. It is suggestive that the holonomy of $T$ indicates the translational displacement of \emph{something} that is parallel transported around the closed curve. However, it is unclear what precisely that something is and how it is parallel transported. It would be self-contradictory if the holonomy of $T$ is naively interpreted as a displacement of \emph{location} experienced by a round trip in some sense of parallel transport, because any round trip, by definition, identifies the final location with the initial location and thus makes no displacement of location. Unlike the case of $R$, the geometric meaning of $T$ in terms of holonomy remains rather obscure.

The goal of this paper is to rigorously derive the parallel between the Riemann curvature and the Cartan torsion from first principles, making precise sense of $\xi$ without any ad hoc prescriptions and obtaining a clear geometric picture of the holonomy of $T$.\footnote{The phrase ``from first principles'' here is used in the ordinary sense, meaning that we make only a few fundamental postulations naturally motivated from physical considerations, once and for all, and everything else (regarding the interplay of $R$ and $T$ through $\xi$) follows deductively. It should be remarked that our work is focused solely on the \emph{kinematical} aspects of MAG, and we leave the \emph{dynamical} aspects for future research. The phrase ``first principles'' used here does not intend to connotate an \textit{ab initio} approach to formulating a dynamical theory of MAG from the action principle. For various \textit{ab initio} approaches starting from the action principle, see e.g.\ \cite{hehl1995metric,Koivisto:2019ejt}. Note that all the approaches studied in the literature so far are essentially based on a conventional vector bundle, instead of an \emph{affine-vector bundle} as proposed in this paper.} Motivated by careful consideration of the Einstein equivalence principle in regard to the Poincar\'{e} symmetry, we propose a new mathematical framework called the \emph{associated affine-vector bundle}, which provides a more suitable arena for the affine group $A(n,\mathbb{R})$ than a conventional associated vector bundle.
Our strategy is to first consider an \emph{Ehresmann connection} endowed on the principal bundle of $A(n,\mathbb{R})$, and then correspondingly define the parallel transport and the covariant derivative on the associated affine-vector bundle in the same spirit of defining the covariant derivative on an associated vector bundle.
The Ehresmann-connection approach is rather formal and less familiar to physicists, but it is advantageous for our purpose of generalizing the notion of parallel transport to the associated affine-vector bundle, as it gives a clear geometric picture of connection independent of local gauge choice.
We obtain the desired results: The field $\xi$ arises naturally in the associated affine-vector bundle as an arbitrary gauge choice of a \emph{reference point}, and the Cartan torsion $T$ and the Reimann curvature $R$ exactly correspond to the holonomies appearing in the ``affine'' part and the ``vector'' part, respectively, of the associated affine-vector bundle.

This paper is organized as follows. In \secref{sec:MAG}, we give a brief overview of the mathematical foundations of MAG. In \secref{sec:Poincare symmetry}, we consider the local Poincar\'{e} symmetry in depth from the perspective of the Einstein equivalence principle, which motivates us to construct the associated affine-vector bundle. In \secref{sec:associated affine-vector bundle}, the associated affine-vector bundle is rigorously formulated. In \secref{sec:exterior covariant derivative} and \secref{sec:curvature}, we then rigorously derive the covariant derivative and the corresponding curvature appearing on the associated affine-vector bundle. In \secref{sec:observational consequences}, we consider possible observational consequences of the holonomy of the Cartan torsion. Finally, the results are summarised and discussed in \secref{sec:summary}.

For the theory of MAG, we follow closely the line of \cite{hehl1995metric}. For the formulation of connections on fiber bundles, we follow closely the line of \cite{Nakahara:2003nw}. Readers who are unfamiliar with the formal quotient-space construction of an associated vector bundle or the formal definition of parallel transport via an Ehresmann connection are advised to read Chapters 9 and 10 of \cite{Nakahara:2003nw} first. Throughout this paper, Latin letters $i,j,\dots$ are used as internal indices for algebras or vectors on fibers, while Greek letters $\mu,\nu,\dots$ are used as external indices for spacetime.\footnote{This convention is the same as that adopted in \cite{blagojevic2002gravitation} but opposite to that in \cite{mielke2017geometrodynamics,gronwald1996gauge,hehl1995metric}.}

\section{Mathematical foundations of metric-affine gauge theory}\label{sec:MAG}
This section gives a brief overview of the mathematical foundations of MAG, following the line of \cite{hehl1995metric}. The main purpose is to introduce basic ideas and define notations for later use.
As a secondary goal, we also endeavor to present these materials with the full rigor so that various confusions resulting from subtleties can be avoided.

In the approach of MAG for an $n$-dimensional spacetime, the gauge group is taken to be the affine group $A(n,\mathbb{R}):=\mathbb{R}^n\rtimes GL(n,\mathbb{R})$, i.e.\ the semidirect product of the vector space $\mathbb{R}^n$ and the degree-$n$ general linear group $GL(n,\mathbb{R})$.
The Lie algebra $\mathfrak{a}(n,\mathbb{R})$ associated with $A(n,\mathbb{R})$ is given by the generators $P_{i}$ of $n$-dimensional translations and the generators ${L^{i}}_{j}$ of $n$-dimensional linear transformations, which satisfy the Lie brackets:
\begin{subequations}\label{algebra}
\begin{eqnarray}
  \label{algebra a}
  \left[P_{i} ,P_{j} \right] &=& 0, \\
  \label{algebra b}
  \left[{L^{{i}}}_{{j}},P_{k}  \right] &=&\delta^{{i}}_{{k}} P_{j},  \\
  \label{algebra c}
  \left[{L^{{i}}}_{{j}},{L^{{m}}}_{{n}} \right] &=&\delta^{{i}}_{{n}}{L^{{m}}}_{{j}}
        -\delta^{{m}}_{{j}}{L^{{i}}}_{{n}},
\end{eqnarray}
\end{subequations}
for ${i},{j},\dots=1,\dots,n$, or ${i},{j},\dots=0,1,\dots,n-1$, depending on the index convention.
The subalgebra spanned by $\{{L^{i}}_{j}\}$ is $\mathfrak{gl}(n,\mathbb{R})$, the Lie algebra associated with $GL(n,\mathbb{R})$. On the other hand, the subalgebra spanned by $\{P^{i}\}$ is $\mathfrak{t}(n,\mathbb{R})$, the Lie algebra associated with the $n$-dimensional translation group $T(n,\mathbb{R})$.\footnote{In the literature, $T(n,\mathbb{R})$ is often denoted as $\mathbb{R}^n$. In this paper, we rigorously distinguish between the translation group $T(n,\mathbb{R})$ and the defining module $\mathbb{R}^n$ of $T(n,\mathbb{R})$ or $GL(n,\mathbb{R})$.}

The Lie algebra $\mathfrak{a}(n,\mathbb{R})$ admits the M\"{o}bius representation $\rho_M$ in the $(n+1)\times(n+1)$ matrix form, which reads as
\begin{equation}
\rho_M({a_{i}}^{j}{L^{{i}}}_{j} +b^{k}P_{k})
=
\left(
\begin{array}{cc}
{a_{i}}^{j}\rho_{n\times n}({L^{{i}}}_{j}) & b^{k}\rho_n(P_{k}) \\
0 & 0 \\
\end{array}
\right)
\equiv
\left(
\begin{array}{cc}
{a_{i}}^{j}{\mbox{$\bar{L}$}^{i}}_{j} & b^{k}\bar{e}_{k} \\
0 & 0 \\
\end{array}
\right),
\end{equation}
where $\rho_{n\times n}:\mathfrak{gl}(n,\mathbb{R})\subset\mathfrak{a}(n,\mathbb{R}) \rightarrow M_{n\times n}(\mathbb{R})$ is the ``identity map'', which maps ${L^{{i}}}_{j}$ to the same $n\times n$ matrix, and $\rho_n:\mathfrak{t}(n,\mathbb{R})\subset\mathfrak{a}(n,\mathbb{R}) \rightarrow \mathbb{R}^n$ is the ``position map'', which maps $P_{k}$ to the $n$-dimensional basis vector $\bar{e}_{k}\in\mathbb{R}^n$; more precisely,\footnote{In this paper, we use barred notations to denote $M_{n\times n}(\mathbb{R})$-valued or $\mathbb{R}^n$-valued objects and unbarred notations for the corresponding $\mathfrak{gl}(n,\mathbb{R})$-valued or $\mathfrak{t}(n,\mathbb{R})$-valued objects. More precisely, for $\eta=\eta^{i}P{i}\in\mathfrak{t}(n,\mathbb{R})\otimes\Omega^p(M)$ and $\kappa={\kappa_{i}}^{j}{L^{i}}_{j} \in\mathfrak{gl}(n,\mathbb{R})\otimes\Omega^p(M)$, where $\Omega^p(M)$ is the space of $p$-forms over $M$, we denote $\bar{\eta}:=\rho_n(\eta)=\eta^{i}\bar{e}_{i} \in\mathbb{R}^n\otimes\Omega^p(M)$ and $\bar{\kappa}:=\rho_{n\times n}(\kappa)={\kappa_{i}}^{j} {\mbox{$\bar{L}$}^{i}}_{j} \in M_{n\times n}(\mathbb{R})\otimes\Omega^p(M)$.}
\begin{subequations}\label{identity maps}
\begin{eqnarray}
{\left(\rho_{n\times n}({L^{{i}}}_{j})\right)^l}_m
&\equiv& {({\mbox{$\bar{L}$}^{i}}_{j})^l}_m
=\delta^{i}_m \delta ^l_{j}, \\
\left(\rho_n(P_{k})\right)^m &\equiv& (\bar{e}_k)^m = \delta^m_{k}.
\end{eqnarray}
\end{subequations}
It is straightforward to prove from \eqref{algebra} that $\rho_M$ is indeed a linear representation of $\mathfrak{a}(n,\mathbb{R})$, i.e.,
\begin{equation}
\rho_M([\alpha,\beta])=[\rho_M(\alpha),\rho_M(\beta)],
\end{equation}
for any $\alpha,\beta\in\mathfrak{a}(n,\mathbb{R})$.

Because $\mathbb{R}^n$ is the defining module of $GL(n,\mathbb{R})$, the Lie algebra $\mathfrak{gl}(n,\mathbb{R})$ naturally acts on $\mathbb{R}^n$ in the matrix form:
\begin{equation}
\big(\rho_{n\times n}({{L}^{i}}_{j})\,\bar{e}_{k}\big)^l
\equiv({\mbox{$\bar{L}$}^{i}}_{j}\,\bar{e}_{k})^l
= {({\mbox{$\bar{L}$}^{i}}_{j})^l}_m (\bar{e}_{k})^m =\delta^l_{j}\delta^{i}_{k}
=\left(\delta^{i}_{k}\bar{e}_{j}\right)^l,
\end{equation}
where \eqref{identity maps} has been used. That is, $\mathbb{R}^n$ is the carrier space of the fundamental representation of $\mathfrak{gl}(n,\mathbb{R})$:
\begin{equation}\label{L on e}
\rho_{n\times n}({L^{i}}_{j})\,\bar{e}_{k}
\equiv {\mbox{$\bar{L}$}^{i}}_{j}\,\bar{e}_{k}
=\delta^{i}_{k}\bar{e}_{j}.
\end{equation}
Meanwhile, the Lie algebra $\mathfrak{t}(n,\mathbb{R})$ also provides a carrier space of the following representation of $\mathfrak{gl}(n,\mathbb{R})$:
\begin{equation}\label{L on P}
\rho_{\mathfrak{t}^n}({L^{i}}_{j})P_{k} := [{L^{i}}_{j},P_{k}]
=\delta^{i}_{k}P_{j}
\end{equation}
according to \eqref{algebra b}. This draws a parallel between $\mathbb{R}^n$ and $\mathfrak{t}^n\equiv\mathfrak{t}(n,\mathbb{R})$ that they are isomorphic to and thus interchangeable with each other as far as their transformations under $\mathfrak{gl}(n,\mathbb{R})$ are concerned.\footnote{\label{foot:subtlety}This point is rarely emphasized and clarified in the literature of MAG. For example, what is intended to be $\rho_{n\times n}({L^{i}}_{j})\rho_n(P_{k})
\equiv {\mbox{$\bar{L}$}^{i}}_{j}\,\bar{e}_{k}$ or $\rho_{\mathfrak{t}^n}({L^{i}}_{j})P_{k} \equiv[{L^{i}}_{j},P_{k}]$ is often denoted in shorthand as ${L^{i}}_{j}P_{k}$ by abuse of notation. The shorthand may look confusing, because for $\alpha,\beta\in\mathfrak{g}$ we have $[\alpha,\beta]\in\mathfrak{g}$ but, rigorously speaking, $\alpha\beta$ is not even well defined.}
More precisely, we have the following commutative diagrams:
\begin{equation}
\begin{tikzcd}
\mathfrak{t}^n\equiv\mathfrak{t}(n,\mathbb{R}) \arrow{d}[left]{\rho_{\mathfrak{t}^n}(\lambda)=[\lambda,\,\cdot\,]} \arrow{r}{\rho_n}
& \mathbb{R}^n \arrow{d}{\rho_{n\times n}(\lambda)\equiv\bar{\lambda}} \\
\mathfrak{t}^n\equiv\mathfrak{t}(n,\mathbb{R}) \arrow{r}{\rho_n}
& \mathbb{R}^n
\end{tikzcd},
\end{equation}
and correspondingly
\begin{equation}
\begin{tikzcd}
\mathfrak{t}^n\equiv\mathfrak{t}(n,\mathbb{R}) \arrow{d}[left]{\mathcal{R}_{\mathfrak{t}^n}(\Lambda)=\Lambda(\,\cdot\,)\Lambda^{-1}} \arrow{r}{\rho_n}
& \mathbb{R}^n \arrow{d}{\rho_{n\times n}(\Lambda)\equiv\bar{\Lambda}} \\
\mathfrak{t}^n\equiv\mathfrak{t}(n,\mathbb{R}) \arrow{r}{\rho_n}
& \mathbb{R}^n
\end{tikzcd},
\end{equation}
where $\Lambda\equiv e^\lambda = e^{({\lambda_{i}}^{j}{L^{i}}_{j})} \in GL(n,\mathbb{R})$ and the representation $\mathcal{R}_{\mathfrak{t}^n}$ of $GL(n,\mathbb{R})$ acting on $\mathfrak{t}(n,\mathbb{R})$ is given by the exponential of $\rho_{\mathfrak{t}^n}$, i.e.,
\begin{equation}
\mathcal{R}_{\mathfrak{t}^n}(\Lambda)P_{i}
:= \Lambda P_{i} \Lambda^{-1}
\equiv P_{i} + [\lambda,P_{i}]+\frac{1}{2!}[\lambda,[\lambda,P_{i}]]+\dots\quad
\end{equation}

Any element $g\in A(n,\mathbb{R})$ can be specified by the two variables $\Lambda\in GL(n,\mathbb{R})$ and $\tau=\tau^{i}P_{i}\in\mathfrak{t}(n,\mathbb{R})$ as\footnote{It seems more elegant to use a pair of algebra-valued variables, $\lambda={\lambda_i}^j{L^i}_j\in \mathfrak{gl}(n,\mathbb{R})$ and $\tau=\tau^{i}P_{i}\in\mathfrak{t}(n,\mathbb{R})$, to specify an element $g\in A(n,\mathbb{R})$ as $g = e^\tau e^\lambda$. But it turns out more convenient to use $\Lambda\in GL(n,\mathbb{R})$ and $\tau\in\mathfrak{t}(n,\mathbb{R})$ instead.}
\begin{subequations}\label{g and inverse}
\begin{eqnarray}
g(\Lambda,\tau) &=& e^\tau\, \Lambda = e^{\tau^{i}\!P_{i}}\,\Lambda, \\
g(\Lambda,\tau)^{-1} &=& \Lambda^{-1} e^{-\tau}= \Lambda^{-1}e^{-\tau^{i}\!P_{i}},
\end{eqnarray}
\end{subequations}
where $ e^\tau\in T(n,\mathbb{R})\subset A(n,\mathbb{R})$ is a translation by the vector $\bar{\tau}=\tau^{i}\bar{e}_{i}\in\mathbb{R}^n$ and $\Lambda\equiv e^{\lambda}\in GL(n,\mathbb{R})\subset A(n,\mathbb{R})$ is a linear transformation exponentiated by $\lambda\in\mathfrak{gl}(n,\mathbb{R})$.
The M\"{o}bius representation $\rho_M$ of $g(\tau,\Lambda)$ then takes the form:
\begin{equation}
\rho_M(g)= \rho_M(e^\tau)\rho_M(e^\lambda) = e^{\rho_M(\tau)} e^{\rho_M(\lambda)}
        =\left(
           \begin{array}{cc}
             \bar{\Lambda} & \bar{\tau} \\
             0 & 1 \\
           \end{array}
         \right),
\end{equation}
where
\begin{subequations}
\begin{eqnarray}
e^{\rho_M(\tau)} &:=& \sum_{n=1}^\infty\frac{1}{n!}\left(
           \begin{array}{cc}
             0 & \bar{\tau} \\
             0 & 0 \\
           \end{array}
         \right)^n
=\left(
           \begin{array}{cc}
             1_{n\times n} & \bar{\tau} \\
             0 & 1 \\
           \end{array}
         \right)
\equiv 1 + \rho_M(\tau), \\
e^{\rho_M(\lambda)} &:=& \sum_{n=1}^\infty\frac{1}{n!}\left(
           \begin{array}{cc}
             \bar{\lambda} & 0 \\
             0 & 0 \\
           \end{array}
         \right)^n
=\left(
           \begin{array}{cc}
             e^{\bar{\lambda}} & 0 \\
             0 & 1 \\
           \end{array}
         \right)
\equiv \left(
           \begin{array}{cc}
             \bar{\Lambda} & 0 \\
             0 & 1 \\
           \end{array}
         \right).
\end{eqnarray}
\end{subequations}
Similarly, we have
\begin{equation}
\rho_M\left(g(\Lambda,\tau)^{-1}\right)=\left(
           \begin{array}{cc}
             \bar{\Lambda}^{-1} & -\bar{\Lambda}^{-1}\bar{\tau}\\
             0 & 1 \\
           \end{array}
         \right),
\end{equation}
which implies
\begin{eqnarray}
g(\Lambda,\tau)^{-1}&=&g(\Lambda^{-1},-\mathcal{R}_{\mathfrak{t}^n}(\Lambda)^{-1}\tau) =g(\Lambda^{-1},-\Lambda^{-1}\tau\Lambda),
\end{eqnarray}
and
\begin{equation}
\rho_M(\Lambda\, e^{\tau}\Lambda^{-1}) =\left(
           \begin{array}{cc}
             1_{n\times n} & \bar{\Lambda}\bar{\tau}\\
             0 & 1 \\
           \end{array}
         \right),
\end{equation}
which implies
\begin{equation}
\Lambda\, e^\tau \Lambda^{-1}
= e^{(\mathcal{R}_{\mathfrak{t}^n}\tau)}
=e^{(\Lambda\tau\Lambda^{-1})}.
\end{equation}
The affine group $A(n,\mathbb{R})$ naturally acts on \emph{affine vectors} $\bar{x}=x^{i}\bar{e}_{i}\in\mathbb{R}^n$ in the following affine transformation:
\begin{equation}\label{affine x}
\bar{x}\rightarrow \bar{x}'=\rho_M(g) \bar{x} = \bar{\Lambda} \bar{x} + \bar{\tau},
\end{equation}
which can be cast in the M\"{o}bius matrix form as
\begin{equation}\label{Mobius}
\left(
\begin{array}{c}
  \bar{x}' \\
  1
\end{array}
\right)
=
\left(
\begin{array}{cc}
\bar{\Lambda} & \bar{\tau} \\
0 & 1 \\
\end{array}
\right)
\left(
\begin{array}{c}
  \bar{x}\\
  1
\end{array}
\right).
\end{equation}

To study the local affine symmetry on a manifold $M$ in the Yang-Mills gauge approach, we locally introduce the affine connection $\mathcal{A}\in \mathfrak{a}(n,\mathbb{R})\otimes \Omega(M)$ as
\begin{equation}
\mathcal{A} = \Gamma^{(L)} + \Gamma^{(T)}
= {L^{i}}_{j} \,\Gamma_{i}^{(L){j}}
+ P_{i}\, \Gamma^{(T){i}}
= {L^{i}}_{j} \,\Gamma_{{i}\mu}^{(L){j}}dx^\mu
+ P_{i}\, \Gamma^{(T){i}}_\mu dx^\mu,
\end{equation}
which in the M\"{o}bius representation reads as
\begin{equation}\label{A in Mobius}
  \bar{\mathcal{A}}:=\rho_M(\mathcal{A})
    =\left(
           \begin{array}{cc}
             \bar{\Gamma}^{(L)} & \bar{\Gamma}^{(T)}\\
             0 & 0 \\
           \end{array}
         \right)
    =\left(
           \begin{array}{cc}
              {\mbox{$\bar{L}$}^{i}}_{j}\,\Gamma_{i}^{(L){j}} & \bar{e}_{i}\, \Gamma^{(T){i}}\\
             0 & 0 \\
           \end{array}
         \right).
\end{equation}
Here, $\Gamma_{i}^{(L){j}}$ and $\Gamma^{(T){i}}$ are the coefficients of the $\mathfrak{gl}(n,\mathbb{R})$-valued one-form $\Gamma^{(L)}$ in the basis ${L^{i}}_{j}$ and of the $\mathfrak{t}(n,\mathbb{R})$-valued one-form $\Gamma^{(T)}$ in the basis $P_{i}$, respectively. These coefficients are by themselves one-forms and thus can be recast as $\Gamma_{i}^{(L){j}}=\Gamma_{{i}\mu}^{(L){j}}dx^\mu$ and $\Gamma^{(T){i}}=\Gamma^{(T){i}}_\mu dx^\mu$.
The connection one-form $\mathcal{A}$ transforms inhomogeneously under the affine gauge transformation $g^{-1}(x)$ as:
\begin{equation}\label{A' and A}
  \mathcal{A}'(x)=g^{-1}(x)\mathcal{A}(x)g(x)+ g^{-1}(x)dg(x),
\end{equation}
The affine connection $\mathcal{A}$ can be viewed to have two parts: the ``linear'' part $\Gamma^{(L)}\in\mathfrak{gl}(n,\mathbb{R})\otimes\Omega(M)$ and the ``translational'' part $\Gamma^{(T)}\in\mathfrak{t}(n,\mathbb{R})\otimes\Omega(M)$.
The gauge transformation rules of them can be easily obtained in the M\"{o}bius representation as
\begin{subequations}\label{gauge transform 1}
\begin{eqnarray}
\label{gauge transform 1a}
  \bar{\Gamma}'^{(L)}
  &=& \bar{\Lambda}^{-1}\bar{\Gamma}^{(L)} \bar{\Lambda} + \bar{\Lambda}^{-1}d\bar{\Lambda},\\
\label{gauge transform 1b}
  \bar{\Gamma}'^{(T)}
  &=& \bar{\Lambda}^{-1} \bar{\Gamma}^{(T)} + \bar{\Lambda}^{-1}\left(d\bar{\tau} + \bar{\Gamma}^{(L)}\bar{\tau}\right)
  \equiv \bar{\Lambda}^{-1}\bar{\Gamma}^{(T)} +\bar{\Lambda}^{-1} \bar{D}^{(L)}\bar{\tau}.
\end{eqnarray}
\end{subequations}
Alternatively, substituting \eqref{g and inverse} and $\mathcal{A}=\Gamma^{(L)}+\Gamma^{(T)}$ into \eqref{A' and A}, by virtue of \eqref{algebra}, we also arrive at the equivalent form:\footnote{Particularly, we have
\begin{equation*}
e^{-\tau}\Gamma^{(L)}e^\tau
= \Gamma^{(L)}+ \frac{(-1)}{1!} [\tau, \Gamma^{(L)}] + \frac{(-1)^2}{2!}[\tau, [\tau, \Gamma^{(L)}]] +\dots
= \Gamma^{(L)} + [\Gamma^{(L)},\tau],
\end{equation*}
and
\begin{equation*}
\Lambda^{-1}e^{-\tau}d\left(e^\tau\Lambda\right) = \Lambda^{-1}d\tau\, \Lambda +\Lambda^{-1}d\Lambda.
\end{equation*}
}
\begin{subequations}\label{gauge transform 2}
\begin{eqnarray}
\label{gauge transform 2a}
  \Gamma'^{(L)} &=& \Lambda^{-1}\Gamma^{(L)} \Lambda+\Lambda^{-1}d\Lambda,\\
\label{gauge transform 2b}
  \Gamma'^{(T)} &=& \Lambda^{-1}\Gamma^{(T)}\Lambda +\Lambda^{-1}\big(d\tau+[\Gamma^{(L)},\tau]\big)\Lambda
  \equiv \Lambda^{-1}\Gamma^{(T)}\Lambda + \Lambda^{-1}\big(D^{(L)}\tau\big)\Lambda \nonumber\\
  &\equiv& \mathcal{R}_{\mathfrak{t}^n}(\Lambda)^{-1}\Gamma^{(T)} + \mathcal{R}_{\mathfrak{t}^n}(\Lambda)^{-1}D_{\mathfrak{t}^n}^{(L)}\tau.\qquad
\end{eqnarray}
\end{subequations}
Here, the exterior covariant derivative with respect to $\Gamma^{(L)}$ is defined as
\begin{equation}\label{DL}
D^{(L)} \equiv D^{(L)}_{\mathfrak{t}^n}\eta := d\eta+[\Gamma^{(L)},\eta]
\equiv d\eta+ \rho_{\mathfrak{t}^n}\big(\Gamma^{(L)}\big)\wedge\eta,
\end{equation}
which maps $\eta\in\mathfrak{t}(n,\mathbb{R})\otimes\Omega^p(M)$ to $D^{(L)}\eta\in\mathfrak{t}(n,\mathbb{R})\otimes\Omega^{p+1}(M)$.\footnote{Let $\zeta=T_{i}\zeta^{i}$ be a $\mathfrak{g}$-valued $p$-form and $\eta=T_{i}\eta^{i}$ be a $\mathfrak{g}$-valued $q$-form. The commutator of them is defined as
\begin{equation*}
[\zeta,\eta] \equiv \zeta\wedge\eta-(-1)^{pq}\eta\wedge\zeta
:=T_{i}T_{j}\,\zeta^{i}\wedge\eta^{j}
-(-1)^{pq}T_{j}T_{i}\,\eta^{j}\wedge\zeta^{i}
=[T_{i},T_{j}]\,\zeta^{i}\wedge\eta^{j},
\end{equation*}
which is a $\mathfrak{g}$-valued $(p+q)$-form.
In case $\zeta=\eta$ and $p=q$ is odd, we have
\begin{equation*}
[\zeta,\zeta]=2\,\zeta\wedge\zeta.
\end{equation*}
In other cases, $\zeta\wedge\eta$ is not necessarily $\mathfrak{g}$-valued in general.
See Section 10.3.2 in \cite{Nakahara:2003nw}.}
For any linear representation $\rho_V$ of $\mathfrak{gl}(n\mathbb{R})$ with $V$ being the carrier space, the exterior covariant derivative associated with $\Gamma^{(L)}$ can also be generalized for a $V$-valued form $v\in V\otimes\Omega^p(M)$, defined as
\begin{equation}\label{DL'}
D^{(L)}_{V}v:=dv+\rho_V\big(\Gamma^{(L)}\big)\wedge v.
\end{equation}
Particularly, we define the shorthand
\begin{equation}\label{DL bar}
\bar{D}^{(L)}:= D_{\mathbb{R}^n}^{(L)}.
\end{equation}

The corresponding affine gauge curvature is defined as
\begin{equation}\label{cal R}
 \mathcal{R}
 :=d\mathcal{A}+\frac{1}{2}[\mathcal{A},\mathcal{A}] \equiv d\mathcal{A}+\mathcal{A}\wedge\mathcal{A}
  = R^{(L)}+ R^{(T)},
\end{equation}
which is again separated into the linear part $R^{(L)}\in\mathfrak{gl}(n,\mathbb{R})\otimes\Omega^2(M)$ and the translational part $R^{(T)}\in\mathfrak{t}(n,\mathbb{R})\otimes\Omega^2(M)$:
\begin{subequations}\label{RL and RT}
\begin{eqnarray}
\label{RL}
  R^{(L)} &\equiv& {L^{{i}}}_{j}R_{i}^{(L){j}}
          = d\Gamma^{(L)} + \frac{1}{2}[\Gamma^{(L)},\Gamma^{(L)}]
          \equiv d\Gamma^{(L)} + \Gamma^{(L)}\wedge\Gamma^{(L)}\nonumber\\
          &=& {L^{{i}}}_{j}\,d\Gamma^{(L){j}}_{i}
             +\frac{1}{2}\left[{L^{{i}}}_{j},{L^{{m}}}_{n} \right]
              \Gamma^{(L){j}}_{i}\wedge\Gamma^{(L){n}}_{m}
          = {L^{{i}}}_{j} \left(d\Gamma^{(L){j}}_{i} +  \Gamma^{(L){j}}_{k}\wedge\Gamma^{(L){k}}_{i}\right), \\
\label{RT}
  R^{(T)} &\equiv& P_{i}R^{(T){i}}
          = d\Gamma^{(T)} + [\Gamma^{(L)},\Gamma^{(T)}]
          \equiv D^{(L)}\Gamma^{(T)} \nonumber\\
          &=&P_{i}\, d\Gamma^{(T){i}}
             + \left[{L^{{i}}}_{{j}},P_{m}  \right]
             \Gamma^{(L){j}}_{{i}}\wedge\Gamma^{(T){m}}
          = P_{i}\left( d\Gamma^{(T){i}}
             + \Gamma^{(L){i}}_{{j}}\wedge\Gamma^{(T){j}} \right).
\end{eqnarray}
\end{subequations}

Compared with \eqref{T and R part b}, the $\mathfrak{gl}(n,\mathbb{R})$-valued two-form $R^{(L)}$ in \eqref{RL} is the familiar Riemann curvature tensor $R=D^{(L)}\Gamma^{(L)}$.
Correspondingly, compared with \eqref{T and R part a}, it is attempting to identity the $\mathfrak{t}(n,\mathbb{R})$-valued two-form $R^{(T)}$ in \eqref{RT} as the familiar Cartan torsion tensor $T=D^{(L)}\theta$, if we identity the $\mathfrak{t}(n,\mathbb{R})$ connection $\Gamma^{(T)}$ as the \emph{coframe} one-form $\theta=\theta^{i}P_{i}\in \mathfrak{t}(n,\mathbb{R})\otimes\Omega(M)$. Unfortunately, because of the inhomogeneous term $\bar{\Lambda}^{-1}\bar{D}^{(L)}\bar{\tau}$, or equivalently $\Lambda^{-1}(D^{(L)}\tau)\Lambda$, appearing in \eqref{gauge transform 1b} and \eqref{gauge transform 2b}, $\Gamma^{(T)}$ \emph{cannot} be identified as $\theta$, which under the gauge transformation $g^{-1}(x)$ transforms as a \emph{vector}, i.e.\footnote{Do not confuse a \emph{vector} with an \emph{affine vector}. The former is insensitive to translation, while the latter transforms as \eqref{affine x} under the gauge transformation $g(x)$.}
\begin{equation}\label{theta as a vector}
\bar{\theta}\stackrel{g^{-1}(x)}{\longrightarrow}\bar{\theta}' = \bar{\Lambda}^{-1}\bar{\theta}.
\end{equation}

Following the suggestion by Trautman \cite{trautman1973structure}, one can introduce an \emph{affine-vector}-valued zero-from $\bar{\xi}=\xi^{i}\bar{e}_{i} \in\mathbb{R}^n\otimes\Omega^0(M)$, which transforms as
\begin{equation}
\bar{\xi}\stackrel{g^{-1}(x)}{\longrightarrow}\bar{\xi}' = \bar{\Lambda}^{-1}(\bar{\xi}-\bar{\tau})
\end{equation}
in accordance with \eqref{affine x},
and define the new one-form as \eqref{theta and xi}, i.e.,
\begin{equation}\label{theta and xi again}
  \theta:=\Gamma^{(T)}+D^{(L)}\xi.
\end{equation}
It can be easily verified that, the transformation law of the one-form $\theta$ defined above just takes the form of \eqref{theta as a vector} and therefore $\theta$ can be identified as the coframe one-form.
Correspondingly, the Cartan torsion $T$ is related to $R^{(T)}$ via \eqref{T and RT}, i.e.,
\begin{equation}
  \bar{T}=\bar{D}^{(L)}\bar{\theta}=\bar{R}^{(T)}+\bar{R}^{(L)}\bar{\xi},
\end{equation}
or equivalently
\begin{equation}
  T=D^{(L)}\theta=R^{(T)}+[R^{(L)},\xi].
\end{equation}

However, as commented in \secref{sec:introduction}, the geometric and physical meaning of $\xi$ remains obscure. In the rest of this paper, we first study the Poincar\'{e} symmetry in more depth and then formulate the new framework of an affine-vector bundle to resolve the problem regarding $\xi$.

\section{Poincar\'{e} symmetry in view of the Einstein equivalence principle}\label{sec:Poincare symmetry}
The \emph{Einstein equivalence principle} states that (Ch.~16 of \cite{Misner:1974qy}): ``\emph{In any and every local Lorentz frame, anywhere and anytime in the universe, all the (nongravitational) laws of physics must take on their familiar special-relativistic forms.}''
However, on the passage from the special-relativistic laws to general-relativistic counterparts by applying the standard ``comma-goes-to-semicolon'' rule \cite{Misner:1974qy}, the resulting covariant laws respect the Lorentz gauge symmetry as well as the symmetry of diffeomorphism, but seem to render irrelevant the translational part of the Poincar\'{e} symmetry of special relativity.\footnote{The ``comma-goes-to-semicolon'' rule is also known as the minimal coupling prescription in Riemannian spacetimes. This prescription has also been studied for the cases that the covariant derivatives are not Riemannian ones. Particularly, it was recently argued that in general one cannot make sense of the notion of minimal coupling that is well defined for arbitrary cases \cite{Delhom:2020hkb,jimenez2020coupling}. This problem does not concerns us here, as our purpose is to take the standard minimal coupling prescription as a typical example to address the issue that the translational part of the Poincar\'{e} symmetry is not manifested. Other prescriptions beyond minimal coupling that have been studied in the literature so far generally bear the same issue.}
In this section, we address this issue in depth by considering the Dirac equation and the Dirac Lagrangian density as representative examples.

The Poincar\'{e} group acting on an $n$-dimensional Minkowski (flat) spacetime is given by $P(1,n-1):=\mathbb{R}^n\rtimes SO(1,n-1)$, where $SO(1,n-1)$ as a proper subgroup of $GL(n,\mathbb{R})$ is the Lorentz group.
The Lie algebra $\mathfrak{p}(1,n-1)$ associated with $P(1,n-1)$ is given by the generators $P_{i}$ of $n$-dimensional translations and the generators $M^{ij}$ of $n$-dimensional Lorentz transformations. The generators $M^{ij}\in\mathfrak{so}(1,n-1)$ are given by particular linear superpositions of the generators ${L^i}_j\in\mathfrak{gl}(n,\mathbb{R})$ as
\begin{equation}
M^{ij}\equiv M^{[ij]} := \eta^{ik}{L^j}_k - \eta^{jk}{L^i}_k,
\end{equation}
where $\eta^{ij}$ is the (inverse of the) metric tensor of the $n$-dimensional Minkowski spacetime.
Consequently, by \eqref{algebra}, the Lie algebra of $\mathfrak{p}(1,n-1)$ satisfies the Lie brackets:
\begin{subequations}\label{algebra Poincare}
\begin{eqnarray}
  \label{algebra P a}
  \left[P^{i} ,P^{j} \right] &=& 0, \\
  \label{algebra P b}
  \left[M^{ij},P^{k} \right] &=& \eta^{ik}P^j-\eta^{jk}P^i,  \\
  \label{algebra P c}
  \left[M^{ij},M^{kl} \right] &=& \eta^{ik}M^{jl} - \eta^{il}M^{jk} -\eta^{jk}M^{il} +\eta^{jl}M^{ik},
\end{eqnarray}
\end{subequations}
where $P^i:=\eta^{ij}P_j$.

The Dirac equation in the Minkowski spacetime is given by
\begin{equation}\label{Dirac eq}
\left(i\gamma^{i}\partial_{i}-m\right)\psi(\bar{x}) = 0,
\end{equation}
where $\bar{x}\equiv x^{i}\bar{e}_{i}\in\mathbb{R}^{n}$,  $x^i$ are the Minkowski coordinates, $\partial_{i}:=\partial/{\partial x^\frak{i}}$, and $\gamma^{i}$ are the $N\times N$ gamma matrices (where $N:=2^{\lfloor n/2\rfloor}$), which satisfy
\begin{equation}
\{\gamma^i,\gamma^j\} \equiv \gamma^i\gamma^j+\gamma^j\gamma^i= 2\eta^{ij}.
\end{equation}
In the Dirac spinor representation, the Lorentz generators are represented as
\begin{equation}
\sigma^{ij}\equiv \sigma^{[ij]}:=\frac{i}{4}[\gamma^{i},\gamma^{j}] \equiv \frac{i}{4} \left(\gamma^i\gamma^j-\gamma^j\gamma^i\right).
\end{equation}
One can easily verify that
\begin{equation}
[\gamma^i,\sigma^{jk}] = i\left(\eta^{ij}\gamma^k-\eta^{ik}\gamma^j\right)
= i\left(\eta^{jl}{({\mbox{$\bar{L}$}^{k}}_{l})^i}_m -\eta^{kl}{({\mbox{$\bar{L}$}^{j}}_{l})^i}_m\right)\gamma^m
\equiv i{(\bar{M}^{jk})^i}_m \gamma^m,
\end{equation}
which is just the infinitesimal form of the finite Lorentz transformation law of $\gamma^i$:
\begin{equation}\label{gamma transform}
\Lambda_D^{-1}\gamma^i\Lambda_D = {(\bar{\Lambda})^i}_j\gamma^j,
\end{equation}
where
\begin{equation}
\Lambda=\exp \Big(\frac{1}{2}\omega_{ij}M^{ij}\Big) \in SO(1,n-1)
\end{equation}
is any Lorentz transformation
and
\begin{equation}
\Lambda_D := \exp\Big(-\frac{i}{2}\omega_{ij}\sigma^{ij}\Big)
\end{equation}
is the Dirac spinor representation of $\Lambda$.

Under a Poincar\'{e} transformation $g(\Lambda,\tau)\in P(1,n-1)\subset A(n,\mathbb{R})$, according to \eqref{affine x}, we have
\begin{subequations}\label{x and partial}
\begin{eqnarray}
\bar{x} &\rightarrow& \bar{x}' = \bar{\Lambda}\bar{x}+\bar{\tau}, \\
\partial_i &\rightarrow& \partial'_i = {(\bar{\Lambda}^{-1})^j}_i \partial_j.
\end{eqnarray}
\end{subequations}
Correspondingly, the Dirac spinor field is transformed as $\psi\rightarrow\psi'$ via\footnote{More precisely, the Dirac spinor field $\psi:\bar{x}\in\mathbb{R}^n\mapsto\psi(\bar{x})\in\mathbb{C}^N$ is to be regarded as a \emph{section} of the fiber bundle with $\mathbb{R}^n$ being the base space and $\mathbb{C}^N$ being the fiber. The Poincar\'{e} transformation moves a given section $\psi$ to a new section $\psi':\bar{x}'\in\mathbb{R}^n\mapsto\psi'(\bar{x}')\in\mathbb{C}^N$.}
\begin{equation}
\psi(\bar{x}) \rightarrow \psi'(\bar{x}') = \Lambda_D\psi(\bar{x}(\bar{x}'))
= \Lambda_D \psi(\bar{\Lambda}^{-1}\bar{x}'-\bar{\Lambda}^{-1}\bar{\tau}).
\end{equation}
The Dirac equation is invariant under the Poincar\'{e} transformation in the sense that, if $\psi(\bar{x})$ satisfies \eqref{Dirac eq}, then it implies
\begin{eqnarray}\label{Dirac eq transform}
\left(i\gamma^i\partial'_i-m\right)\psi'(\bar{x}')
&=&\left(i\gamma^i\partial'_i-m\right)
\Lambda_D \psi(\bar{\Lambda}^{-1}\bar{x}'-\bar{\Lambda}^{-1}\bar{\tau}) \nonumber\\
&=& \Lambda_D \left(i\Lambda_D^{-1}\gamma^i\Lambda_D {(\bar{\Lambda}^{-1})^j}_i \partial_j-m\right)\psi(\bar{x}) \nonumber\\
&=& \Lambda_D \left(i{(\bar{\Lambda})^i}_k\gamma^k {(\bar{\Lambda}^{-1})^j}_i \partial_j-m\right)\psi(\bar{x}) \nonumber\\
&=& \Lambda_D \left(i\gamma^j \partial_j-m\right)\psi(\bar{x})
=0,
\end{eqnarray}
where \eqref{gamma transform}, \eqref{x and partial}, and ${(\bar{\Lambda})^i}_k {(\bar{\Lambda}^{-1})^j}_i =\delta^j_k$ (i.e.\ $\bar{\Lambda}^{-1}\bar{\Lambda}=1_{n\times n}$) have been used.

Furthermore, defining
\begin{equation}
\bar{\psi} = \psi^\dag\gamma^0,
\end{equation}
we have the transformation law:
\begin{equation}\label{psi bar transform}
\bar{\psi}(\bar{x}) \rightarrow \bar{\psi}'(\bar{x}') =  \bar{\psi}(\bar{\Lambda}^{-1}\bar{x}'-\bar{\Lambda}^{-1}\bar{\tau}) \Lambda_D^{-1}.
\end{equation}
It follows from \eqref{Dirac eq transform} and \eqref{psi bar transform} that the Dirac lagrangian density defined as
\begin{equation}
\mathcal{L}_\mathrm{Dirac} := \bar{\psi}\left(i\gamma^{i}\partial_{i}-m\right)\psi
\end{equation}
is invariant under the Poincar\'{e} transformation as well.

\begin{figure}

\centering
 \includegraphics[width=0.6\textwidth]{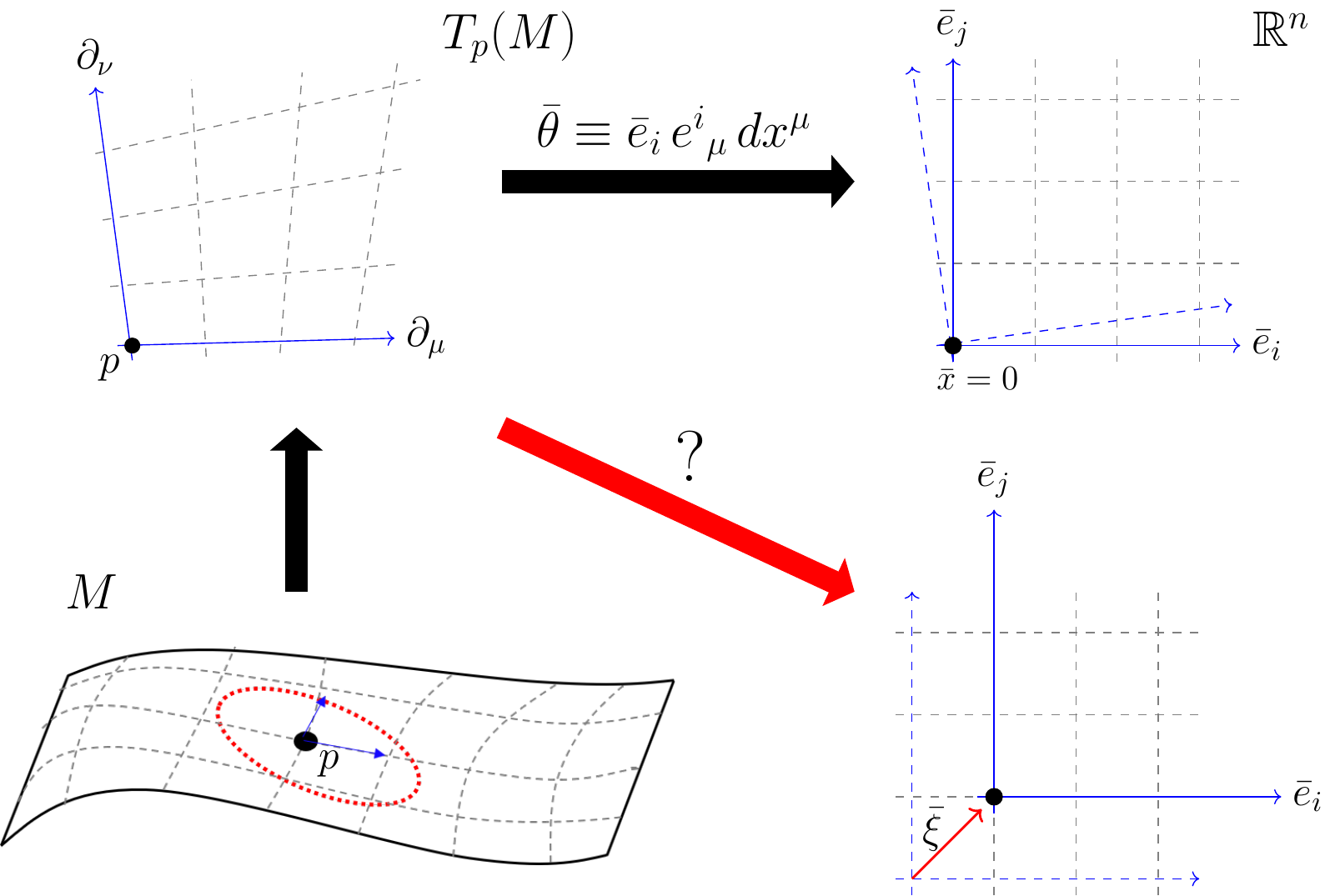}

\caption{A spacetime manifold $M$ (lower left) naturally gives rise to a tangent space $T_p(M)$ (upper left) at any point $p\in M$. $T_p(M)$ is to be identified as the local Lorentz frame $\mathbb{R}^n$ of a freely falling laboratory via the ``soldering form'' $\bar{\theta}\equiv \bar{e}_i{e^i}_\mu dx^\mu$ (see upper right). Within the local frame in the locality of $p$, a local Lorentz transformation rotates or boosts the orthonormal coordinates $x^i$ to $x^{\prime i}$ while leaving the origin unmoved. On the other hand, a local Poincar\'{e} translation displaces the coordinate origin from the point $p$, and consequently the translated local frame can no longer be identified as $T_p(M)$ (see lower right).\label{fig:local frame}}
\end{figure}

The Dirac equation and the Dirac Lagrangian can be generalized from the context of special relativity to that of general relativity. In general relativity, we have an $n$-dimensional manifold $M$ as the curved spacetime, which is locally coordinated by $x^\mu$ (not to be confused with the Minkowski coordinates $x^i$). At each point $p\in M$, we have a tangent space $T_p(M)$, which is isomorphic to $\mathbb{R}^n$ and is to be identified as the local Lorentz frame of a freely falling laboratory (see \figref{fig:local frame}). Consider a Dirac spinor field $\varphi$ on $M$, i.e., $\varphi: p\in M \mapsto \varphi(x^\mu(p)) \in\mathbb{C}^N$. In the immediate vicinity of a given point $p$, the field $\varphi$ \emph{locally} gives rise to a Dirac spinor field $\psi: \bar{x}\in\mathbb{R}^n\cong T_p(M) \mapsto \psi(\bar{x})\in\mathbb{C}^N$ in the locality of the origin $\bar{x}=0$ as measured by a local freely falling laboratory.
The local field $\psi$ in the locality of $\bar{x}=0$ is to be understood as the ``best'' linear approximation of $\varphi$ in the vicinity of $p$; its exact formulation is to be properly prescribed in accordance with the Einstein equivalence principle as follows. Firstly, the value of $\psi$ at the origin $\bar{x}=0$ of $\mathbb{R}^n$ is prescribed to be identical to that of $\varphi$ at $p$, i.e.,
\begin{equation}\label{EP psi 1}
\psi(\bar{x})\big|_{\bar{x}=0} = \varphi(x^\mu)\big|_{x^\mu=x^\mu(p)}.
\end{equation}
Secondly, in order to reflect the local features of $\psi$ around the locality of $\bar{x}=0$ in the local Lorentz frame, we need also to specify the derivatives of $\psi$ with respect to $\bar{x}$. As the Minkowski coordinates $x^i$ of a freely falling frame are supposed to be locally identified as \emph{Riemann normal coordinates}, it is most natural to take the prescription:
\begin{equation}\label{EP psi 2}
(\partial_i\psi)(\bar{x})\big|_{\bar{x}=0}
=e_i(x^\mu)\!\iprod D^{(D)} \varphi(x^\mu)\big|_{x^\mu=x^\mu(p)}
={e_i}^\nu(x^\mu) D^{(D)}_\nu \varphi(x^\mu)\big|_{x^\mu=x^\mu(p)},
\end{equation}
where $D^\mathrm{(D)}$ is the exterior covariant derivative associated with the $SO(1,n-1)$ spin connection, and
\begin{equation}
e_i\equiv{e_i}^\mu\partial_\mu
\end{equation}
are the \emph{frame field} (also called \emph{vielbein}), i.e.\ a set of $n$ orthonormal vector fields.\footnote{The frame field is also known as a \emph{tetrad} or \emph{vierbein} in the case of $n=4$.} The frame field is related to the vector-valued \emph{coframe} one-form field
\begin{equation}\label{coframe}
\bar{\theta} \equiv \bar{e}_i\,{e^i}_\mu dx^\mu
\end{equation}
via
\begin{equation}
{e_i}^\mu{e^j}_\mu = \delta_i^j.
\end{equation}
The coframe field $\bar{\theta}$ provides a mathematical facility that naturally ``solders'' a tangent vector $X=X^\mu\partial_\mu\in T_p(M)$ to $\bar{X}=\bar{\theta}(X)\equiv X\!\iprod\bar{\theta}={e^i}_\mu X^\mu \bar{e}_i\in\mathbb{R}^n$ in the local frame (see \figref{fig:local frame}). Particularly, it maps $e_i\in T_p(M)$ to $\bar{\theta}(e_i)\equiv e_i\!\iprod \bar{\theta}=\bar{e}_i\in\mathbb{R}^n$.
Furthermore, the exterior covariant derivative $D^{(D)}$ is defined as
\begin{equation}
D^{(D)} := d + \rho_D\big(\Gamma^{(SO)}\big)
\equiv d + \Gamma_D^{(SO)}
\equiv \left(\partial_\mu -i\sigma^{ij}\Gamma^{(SO)}_{ij\mu}\right)dx^\mu,
\end{equation}
where $\Gamma^{(SO)}\equiv M^{ij}\Gamma^{(SO)}_{ij}$ is the $\mathfrak{so}(1,n-1)$-valued connection one-form, which is the same as $\Gamma^{(L)}\equiv{L^{i}}_{j} \,\Gamma_{{i}}^{(L){j}}$ except that its algebra value is now restricted to the subalgebra $\mathfrak{so}(1,n-1)$ of $\mathfrak{gl}(n,\mathbb{R})$, and $\rho_D(M^{ij})=-i\sigma^{ij}$ is the Dirac spinor representation of $\mathfrak{so}(1,n-1)$, which maps an element of $\mathfrak{so}(1,n-1)$ to an $N\times N$ matrix.

Under a local Lorentz gauge transformation $\Lambda(x^\mu)\in SO(1,n-1)$, we have
\begin{subequations}\label{SO gauge transform}
\begin{eqnarray}
\label{SO gauge transform a}
  \varphi(x^\mu) &\rightarrow& \varphi'(x^\mu) = \Lambda_D(x^\mu)\varphi(x^\mu), \\
\label{SO gauge transform b}
  e_i(x^\mu) &\rightarrow& e'_i(x^\mu) = {(\bar{\Lambda}(x^\mu)^{-1})^j}_i e_j(x^\mu), \\
\label{SO gauge transform c}
  \Gamma_D^{(SO)}(x^\mu) &\rightarrow& \Gamma_D^{\prime(SO)}(x^\mu) = \Lambda_D(x^\mu) \Gamma_D^{(SO)}(x^\mu) \Lambda_D^{-1}(x^\mu)+ \Lambda_D(x^\mu) d\Lambda_D^{-1}(x^\mu),
\end{eqnarray}
\end{subequations}
where \eqref{SO gauge transform c} is given in accordance with \eqref{gauge transform 2a}.\footnote{Note that \eqref{A' and A} is expressed for the gauge transformation $g^{-1}(x)$, instead of $g(x)$. Thus, we have to interchange $\Lambda$ and $\Lambda^{-1}$ when applying \eqref{gauge transform 2a} here.}
Consequently, these imply
\begin{eqnarray}
D^{(D)}\varphi(x^\mu) &\rightarrow& (D^{(D)}\varphi)'(x^\mu)
\equiv \big(d+\Gamma_D^{\prime(SO)}(x^\mu)\big)\varphi'(x^\mu)= \Lambda_D(x^\mu) D^{(D)}\varphi(x^\mu).
\end{eqnarray}
According to  \eqref{EP psi 1} and \eqref{EP psi 2}, from the viewpoint of a freely falling laboratory whose frame coordinates are given by $\bar{x}=x^i\bar{e}_i\in\mathbb{R}^n$, the local $SO(1,n-1)$ gauge transformation upon the local field reads as
\begin{subequations}
\begin{eqnarray}
\psi(\bar{x})\big|_{\bar{x}=0} &\rightarrow& \psi'(\bar{x}')\big|_{\bar{x}'=0} = \Lambda_D(p) \psi(\bar{x})\big|_{\bar{x}=0},\\
(\partial_i\psi)(\bar{x})\big|_{\bar{x}=0} &\rightarrow& (\partial_i\psi)'(\bar{x}')\big|_{\bar{x}'=0} = {(\bar{\Lambda}(p)^{-1})^j}_i \Lambda_D(p)(\partial_j\psi)(\bar{x})\big|_{\bar{x}=0}.
\end{eqnarray}
\end{subequations}
Consequently, by the similar tricks used for deriving \eqref{Dirac eq transform}, it can be shown that
\begin{equation}
\left(i\gamma^i(\partial_i\psi)'(\bar{x}')-m\psi'(\bar{x}')\right)\big|_{\bar{x}'=0} =\Lambda_D(p) \left(i\gamma^i(\partial_i\psi)(\bar{x})-m\psi(\bar{x})\right)\big|_{\bar{x}=0}.
\end{equation}
That is, in the local Lorentz frame of a freely falling laboratory, if $\psi(\bar{x})$ in the locality of an event $p$ satisfies the Dirac equation with respect to the local Minkowski coordinates $\bar{x}$, then the $SO(1,n-1)$-transformed version $\psi'(\bar{x}')$ in the locality of $p$ satisfies the Dirac equation as well.
We have just affirmed the Einstein equivalence principle with regard to the Lorentz, i.e.\ $SO(1,n-1)$, symmetry of the Dirac equation.
Therefore, the covariant generalization of the Dirac equation is prescribed as
\begin{equation}
\left(i\gamma^{i}{e_i}^\mu D^{(D)}_\mu-m\right)\varphi(x^\mu) = 0,
\end{equation}
which has been shown to respect the local Lorentz gauge symmetry.
Similarly, the covariant generalization of the Dirac Lagrangian density is prescribed as
\begin{equation}
\mathcal{L}_\mathrm{Dirac} := \bar{\varphi}\left(i\gamma^{i}{e_i}^\mu D^{(D)}_\mu-m\right)\varphi,
\end{equation}
which respects the local Lorentz gauge symmetry as well.

Unfortunately, while the covariant Dirac equation and the covariant Dirac Lagrangian density are invariant under Lorentz gauge transformations as discussed above as well as under diffeomorphisms of $x^\mu\rightarrow x'^\mu=x'^\mu(x^\mu)$ (as they are in covariant forms), they are oblivious of the translational part of the Poincar\'{e} symmetry. That is, we can make sense of a local Lorentz transformation corresponding to $\bar{x} \rightarrow \bar{x}' = \bar{\Lambda}(p)\bar{x}$ within the local Lorentz frame, but cannot prescribe a local translation $\bar{x}\rightarrow\bar{x}'=\bar{x}+\bar{\tau}(p)$ within the Lorentz frame as any attempt to do so is meant to spoil the fact that the local Lorentz frame in the locality of $p$ is identified as $T_p(M)$.
This obstacle is not only for the Dirac spinor field but for any generic field that is described as a local section of a vector bundle over $M$ associated with the Lorentz group.\footnote{\label{foot:field as a local section}Particularly, the Dirac field $\varphi(x^\mu)$ can be viewed as a local section of the $\mathbb{C}^N$ vector bundle over $M$ associated with $SO(1,n-1)$, and a scalar field $\phi(x^\mu)$ a local section of the $\mathbb{R}$ or $\mathbb{C}$ vector bundle.}
Lacking an adequate mathematical language to address a local translation suggests that the Einstein equivalence principle has not been completely fulfilled in the standard comma-goes-to-semicolon approach, because special-relativistic laws are supposed to respect the full symmetry of the Poincar\'{e} group, not just its Lorentz subgroup.

In order to make sense of the local Poincar\'{e} transformation, in the next section, we propose a new mathematical framework called the \emph{associated affine-vector bundle}, which extends the structure of an associated vector bundle by rendering the Poincar\'{e} translation as an affine transformation on the fiber. Let $v\in V$ represents a local value of a generic field in the vector fiber $V$. The idea is to augment $v$ with an \emph{affine vector} $\bar{\xi}\in\mathbb{R}^n$ serving as a \emph{reference point}, so that $f=(\bar{\xi},v)\in F=\mathbb{R}^n\times V$ represents the local field with reference to $\bar{\xi}$, which indicates the displacement between the designated origin of the local frame and the point $p$. See \figref{fig:local frame}.

One might argue that designating the reference point is purely a matter of arbitrary gauge choice, so it makes no difference to take its involvement into account. This argument is true only if one considers the local physics that is completely independent of the neighboring local frame. If one considers the covariant derivative of the reference point in the sense of ``parallel transport'', it will yield physical significance in relation to the Cartan torsion. This is the topic we will investigate closely. To keep our analysis as generic as possible, in the following, we will study the local translational transformation in the broader context of the affine group $A(n,\mathbb{R}):=\mathbb{R}^n\rtimes GL(n,\mathbb{R})$, instead of the Poincar\'{e} group proper.

\section{Associated affine-vector bundle}\label{sec:associated affine-vector bundle}
Given a principal bundle $P(M,G)$ whose fiber is identical to the structure group $G$, one can formulate an \emph{associated vector bundle} $P\times_\rho V$ with $V$ being a vector space and $\rho$ being the presentation of $G$ over $V$ via the quotient-space construction (see Sec.~9.4.2 of \cite{Nakahara:2003nw}).
The principal bundle $P(M,G)$ can be endowed with an \emph{Ehresmann connection} one-form $\omega$ and its local form as a pullback to the base space $M$ is the well-known gauge potential (see Sec.~10.1 of \cite{Nakahara:2003nw}).
The connection defined on $P(M,G)$ naturally gives rise to the covariant derivative on the associated vector bundle $P\times_\rho V$ and the resulting covariant derivative involves the same gauge potential (see Sec.~10.4 of \cite{Nakahara:2003nw}).\footnote{\label{foot:vector bundle}Therefore, if we consider parallel transport on a vector bundle associated with $A(n,\mathbb{R})$, we will simply obtain a result identical (or, more precisely, homomorphic) to \eqref{RL and RT}. Without going beyond the framework of an associated vector bundle, we are unable to solve the problem of $\xi$.}

In the approach of MAG, the group $G$ is given by the affine group $A(n,\mathbb{R}):=\mathbb{R}^n\rtimes GL(n,\mathbb{R})$.
To reflect the translational symmetry more faithfully, we develop a slightly different framework by constructing an \emph{associated affine-vector bundle} in the same spirit of defining an associated vector bundle. The associated affine-vector bundle is defined below and denoted as $E=P\times_{\rho_A}F$, where $F=\mathbb{R}^n\times V$ is the affine-vector space and $\rho_A$ is the affine (nonlinear) representation of $A(n,\mathbb{R})$ acting on $F$ defined below.

Let $g=g(\Lambda,\tau)\in G$, where $\Lambda\in GL(n,\mathbb{R})$ and $\tau\in\mathfrak{t}(n,\mathbb{R})$, act on $(u,f)\in P\times F$ as follows:
\begin{eqnarray}
g(u,f)
&\equiv& g(u,(\bar{\xi},v))= (ug,\rho_A(g)^{-1}f) \nonumber\\
&\equiv& \left(ug,\left(\rho_M(g)^{-1}\bar{\xi},\rho_V(\Lambda)^{-1}v\right)\right) \nonumber\\
&\equiv&\left(ug, (\bar{\Lambda}^{-1}(\bar{\xi}-\bar{\tau}),\rho_V(\Lambda)^{-1}v)\right),
\end{eqnarray}
where $f=(\bar{\xi},v)\in F=\mathbb{R}^n\times V$, $\rho_V(\Lambda)$ is the linear representation of $GL(n,\mathbb{R})$ acting on $v\in V$, and $\rho_M(g)$ is the M\"{o}bius representation of $A(n,\mathbb{R})$ acting on affine vectors as defined in \eqref{affine x}.
That is, the ``vector'' part $V$ of $F$ responses only to the $GL(n,\mathbb{R})$ part of $G$, whereas the ``affine'' part $\mathbb{R}^n$ of $F$ serves as a ``reference point'', which transforms as an affine vector.\footnote{Note that $\rho_A(g):F\rightarrow F$ is a representation of $g\in A(n,\mathbb{R})$ in the sense $\rho_A(g_1g_1)=\rho_A(g_1)\rho_A(g_2)$ and $\rho_A(e)f=f$, where $e=g(\Lambda=\mathds{1},\tau=0)$. However, it is a \emph{nonlinear} representation. If $\rho_A(g)f_1=f'_1$ and $\rho_A(g)f_2=f'_2$, it does \emph{not} imply $\rho_A(g)(\alpha f_1+\beta f_2)=\alpha f'_1+\beta f'_2$. By contrast, the vector part $\rho_V(\Lambda)$ is linear.}
The associated affine-vector bundle $E=P\times_{\rho_A} F$ is then defined as the quotient space $(P\times F/{\sim})$ via the following equivalence relation
\begin{equation}\label{equivalence rel}
(u,f) \sim (ug,\rho_A(g)^{-1}f).
\end{equation}
The affine-vector bundle $E$ is a fiber bundle over $M$ with $F$ being the fiber in the sense that its fiber bundle structure is given by the projection, trivialization, and transition functions as defined in the following.

The projection $\pi_E:E\rightarrow M$ is defined as $\pi_E(u,f)=\pi(u)$, where $\pi:P\rightarrow M$ is the projection of $P(M,G)$. The projection $\pi_E$ is well defined under the equivalence relation \eqref{equivalence rel}, since we have $\pi(ug)=\pi(u)$ and it follows $\pi_E(ug,\rho_A(g)^{-1}f)=\pi_E(u,f)$. The local trivialization $\psi_\mathcal{i}:U_\mathcal{i}\times F\rightarrow\pi_E^{-1}(U_\mathcal{i})$ is given by $\psi_\mathcal{i}(p,f)=(\phi_\mathcal{i}(p,e),f)$, where $U_\mathcal{i}\subset M$ is an open set of $M$, $p\in U_\mathcal{i}$ is a point in $M$, $e\in G$ is the identity, and $\phi_\mathcal{i}:U_\mathcal{i}\times G\rightarrow\pi^{-1}(U_\mathcal{i})$ is the trivialization of $P(M,G)$.
It is easy to show that $\pi_E\circ\psi_\mathcal{i}(p,f)=p$.

Let $\phi_\mathcal{i}^{-1}(u)=(p,g_\mathcal{i})$, where $u\in\pi^{-1}(U_\mathcal{i})$ and $p=\pi(u)$. On $U_\mathcal{i}\cap U_\mathcal{j}\neq\emptyset$, the two trivializations $\phi_\mathcal{i}$ and $\phi_\mathcal{j}$ are related by a smooth map $t_\mathcal{ij}:U_\mathcal{i}\cap U_\mathcal{j}\rightarrow G$ via
\begin{equation}
\phi_\mathcal{j}(p,g_\mathcal{j}) = \phi_\mathcal{i}(p,t_\mathcal{ij}(p)g_\mathcal{j}),
\end{equation}
where the maps $t_\mathcal{ij}$ are the transition functions of $P$.
As $P$ is a principal bundle, the right action of $g\in G$ on $\pi^{-1}(U_\mathcal{i})$ is defined as $\phi_\mathcal{i}^{-1}(ug)=(p,g_\mathcal{i}g)$, or equivalently
\begin{equation}\label{right action}
ug = \phi_\mathcal{i}(p,g_\mathcal{i}g).
\end{equation}
This leads to
\begin{equation}
ug = \phi_\mathcal{j}(p, g_\mathcal{j}g) = \phi_\mathcal{i}(p,t_\mathcal{ij}(p)g_\mathcal{j}g) = \phi_\mathcal{i}(p,g_\mathcal{i}g),
\end{equation}
where $g_\mathcal{i}=t_\mathcal{ij}(p)g_\mathcal{j}$. That is, the right action is well defined and independent of local trivializations.
Now, regarding the trivializations of $E$, we have
\begin{eqnarray}
\psi_\mathcal{j}(p,f_\mathcal{j}) &\equiv& (\phi_\mathcal{j}(p,e),f_\mathcal{j})
=(\phi_\mathcal{i}(p,t_\mathcal{ij}(p)),f_\mathcal{j})
=(\phi_\mathcal{i}(p,e)t_\mathcal{ij}(p),f_\mathcal{j}) \nonumber\\
&\sim& (\phi_\mathcal{i}(p,e),\rho_A(t_\mathcal{ij}(p))^{-1}f_\mathcal{j})
\equiv \psi_\mathcal{i}(p,\rho_A(t_\mathcal{ij}(p))^{-1}f_\mathcal{j}),
\end{eqnarray}
where we have used \eqref{equivalence rel} and \eqref{right action}. That is, the transition functions of $E$ are given by $\rho_A(t_\mathcal{ij}(p))^{-1}$, where $t_\mathcal{ij}(p)$ are the transition functions of $P$.

\section{Covariant derivative on the affine-vector bundle}\label{sec:exterior covariant derivative}
If a principle bundle $P$ is endowed with an \emph{Ehresmann connection} one-form $\omega\in\mathfrak{g}\otimes T^*P$, it naturally gives rise to the notion of \emph{parallel transport} as one can define the \emph{horizontal lift} of a given curve $\gamma:[0,1]\rightarrow M$. A curve $\tilde{\gamma}:[0,1]\rightarrow P$ is said to be a horizontal lift of $\gamma$ if $\pi\circ\tilde{\gamma}=\gamma$ and $\omega(\tilde{X})=0$, where $\tilde{X}=\tilde{\gamma}_*(X)$ with $X(t)\in T_{\gamma(t)}M$ being a tangent vector to $\gamma(t)$.

Let $\sigma_\mathcal{i}\in\Gamma(U_\mathcal{i},P):U_\mathcal{i}\rightarrow P$ be an arbitrary local section of $P$ on $U_\mathcal{i}$. The well known gauge potential $\mathcal{A}_\mathcal{i}\in\mathfrak{g}\otimes\Omega(U_\mathcal{i})$ is given as the pullback of $\omega$ via $\sigma_\mathcal{i}$, i.e.
\begin{equation}\label{sigma and A}
\mathcal{A}_\mathcal{i} = \sigma_\mathcal{i}^*\omega.
\end{equation}
As two sections are related by transition functions, i.e., $\sigma_\mathcal{j}(p)=\sigma_\mathcal{i}(p)t_\mathcal{ij}(p)$, it turns out $\mathcal{A}_\mathcal{i}$ follows the well known gauge transformation rule:
\begin{equation}\label{gauge transform}
\mathcal{A}_\mathcal{j} = t_\mathcal{ij}^{-1}\mathcal{A}_\mathcal{i}t_\mathcal{ij}+t_\mathcal{ij}^{-1}dt_\mathcal{ij}.
\end{equation}
Conversely, if gauge potentials that are locally defined for an open covering of $M$ satisfy \eqref{gauge transform}, we can uniquely construct an Ehresmann connection one-form $\omega$ for $P$ from these local gauge potentials.

Given an Ehresmann connection on $P$, we can define the \emph{covariant derivative} on an associated affine-vector bundle $E=P\times_{\rho_A}F$ just in the same spirit of defining the covariant derivative on an associated vector bundle $P\times_\rho V$ (see Sec.~10.4 of \cite{Nakahara:2003nw}).

First, we consider a local section $s_v\in\Gamma(U_\mathcal{i},E)$ of $E$ on $U_\mathcal{i}$, i.e.\ $s_v:U_\mathcal{i}\rightarrow E$, as a representative of the equivalence class associated with the relation \eqref{equivalence rel}:
\begin{equation}\label{s v}
s_v(p) = \left[\sigma_\mathcal{i}(p),\left(\bar{\xi}_\mathcal{i}(p),v(p)\right)\right],
\end{equation}
where $\sigma_\mathcal{i}:U_\mathcal{i}\rightarrow P$ is a local section of $P$ on $U_\mathcal{i}$, $\bar{\xi}_\mathcal{i}:U_\mathcal{i}\rightarrow\mathbb{R}^n$, and $v:U_\mathcal{i}\rightarrow V$.
The arbitrary choice of $\sigma_\mathcal{i}$ amounts to different gauge fixing as indicated in \eqref{sigma and A}. In the affine-vector bundle associated with $G=A(n,\mathbb{R})$, we have the additional arbitrariness of choosing different ``reference points'', which amounts to specifying $\bar{\xi}_\mathcal{i}(p)$.

For a given curve $\gamma(t)$, a section $s_v(\gamma(t))$ along $\gamma(t)$ is said to be parallel transported if, in the representation $s_v(\gamma(t))=\left[(\tilde{\gamma}(t),(\bar{\xi}_\mathcal{i}(\gamma(t)),v(\gamma(t))))\right]$, $\bar{\xi}_\mathcal{i}$ and $v$ remain constant (i.e., independent of $t$). The notion of parallel transport is well defined, since if $\tilde{\gamma}'(t)$ is another horizontal lift of $\gamma(t)$, we have $\tilde{\gamma}'(t)=\tilde{\gamma}(t)g$ for a constant $g\in G$ and consequently
\begin{equation}
[\tilde{\gamma},(\bar{\xi}_\mathcal{i},v)] = [\tilde{\gamma}'g^{-1},(\bar{\xi}_\mathcal{i},v)]
= [\tilde{\gamma}',(\rho_M(g)^{-1}\bar{\xi}_\mathcal{i},\rho_V(g)^{-1}v)]
\equiv [\tilde{\gamma}',(\bar{\xi}'_\mathcal{i},v')],
\end{equation}
which follows that $\bar{\xi}'_\mathcal{i}$ and $v'$ are constant as well.

The notion of parallel transport enables us to define the covariant derivative of a section $s_v(p)$. Let $p=\gamma(t=0)\in M$ and $X\in T_pM$ be a tangent vector to $\gamma$ at $p$, i.e., ${dx^\mu(\gamma(t))}/{dt}\big|_{t=0}=X^\mu$. The covariant derivative of $s(p)$ with respect to $X$ is defined as
\begin{equation}
\nabla_X s_v := \left[\tilde{\gamma}(0),
\left.\frac{d}{dt}\Big(\bar{\xi}_\mathcal{i}(\gamma(t)),v(\gamma(t))\Big)\right|_{t=0}
\right].
\end{equation}
Again, it is easy to show that this definition is well defined, regardless of choosing a different horizontal lift $\tilde{\gamma}'(t)$.
Once we have defined $\nabla_X s$, we can also define the exterior covariant derivative $\nabla:\Gamma(U,E)\rightarrow\Gamma(U,E)\otimes\Omega(U)$, which maps a section $s_v\in\Gamma(U,E)$ to a section-valued one-form, by
\begin{equation}
\nabla s_v(X) := \nabla_X s_v,
\end{equation}
where $X\in\mathscr{X}(U)$ is a vector field over an open set $U$.

So far, the ideas of parallel transport and covariant derivative on the associated affine-vector bundle are the same as those on the associated vector bundle. However, because the section $s_v(p)$ has to specify $\bar{\xi}_\mathcal{i}(p)$ as an extra ``gauge fixing'', the \emph{local} expression for the covariant derivative leads to a crucial difference.

Given a local section $\sigma_\mathcal{i}\in\Gamma(U_\mathcal{i},P)$, a horizontal lift $\tilde{\gamma}(t)$ of $\gamma(t)$ can be expressed as $\tilde{\gamma}(t)=\sigma_\mathcal{i}(t)g_\mathcal{i}(t)$, where $g_\mathcal{i}(t)\equiv g(\Lambda_\mathcal{i}(t),\tau(t)):= g_\mathcal{i}(\gamma(t))\in A(n,\mathbb{R})$.
A local section along $\gamma(t)$ expressed in the form \eqref{s v} then leads to
\begin{equation}
s_{v}(t) := s_{v}(\gamma(t)) = \left[\tilde{\gamma}(t)g_\mathcal{i}(t)^{-1},(\bar{\xi}_\mathcal{i}(t),v(t))\right]
= \left[\tilde{\gamma}(t),\rho_A(g_\mathcal{i}(t))^{-1}(\bar{\xi}_\mathcal{i}(t),v(t))\right],
\end{equation}
where $f(t)\equiv\left(\bar{\xi}_\mathcal{i}(t),v(t)\right):=\left(\bar{\xi}_\mathcal{i}(\gamma(t)),v(t)\right)$.
By the identity ${dg(t)^{-1}}/{dt}=-g(t)^{-1}({dg(t)}/{dt})g(t)^{-1}$, we then have
\begin{eqnarray}\label{exterior covariant derivative}
\nabla_X s_{v} &=& \left[\tilde{\gamma}(0),
\left.\frac{d}{dt}\Big(\rho_A(g_\mathcal{i}(t))^{-1}(\bar{\xi}_\mathcal{i}(t),v(t))\Big)\right|_{t=0}
\right]\nonumber\\
&=& \left[\tilde{\gamma}(0),
\left.-\rho_A(t)^{-1}\frac{d\rho_A(t)}{dt}\rho_A(t)^{-1}f(t)+\rho_A(t)^{-1}\frac{df(t)}{dt}\right|_{t=0}
\right] \nonumber\\
&=&
\left[\tilde{\gamma}(0)g_\mathcal{i}(0)^{-1},
\left.-\frac{d\rho_A(t)}{dt}\rho_A(t)^{-1}f(t)+\frac{df(t)}{dt}\right|_{t=0}
\right] \nonumber\\
&=&\left[\sigma_\mathcal{i}(0),
\left.\left(-\frac{d\rho_M(t)}{dt}\rho_M(t)^{-1}\bar{\xi}_\mathcal{i}(t)+\frac{d\bar{\xi}_\mathcal{i}(t)}{dt},
-\frac{d\rho_V(t)}{dt}\rho_V(t)^{-1}v(t)+\frac{dv(t)}{dt}\right)\right|_{t=0}
\right],\nonumber\\
\end{eqnarray}
where $\rho_A(t):=\rho_A(g_\mathcal{i}(t))$, $\rho_M(t):=\rho_M(g_\mathcal{i}(t))$, and $\rho_V(t):=\rho_V(\Lambda_\mathcal{i}(t))$.
As $\sigma_\mathcal{i}$ is a section of $P(M,G)$ and $\tilde{\gamma}(t)=\sigma_\mathcal{i}(\gamma(t))g_\mathcal{i}(t)$, it can be shown that (see Eq.~10.13 in \cite{Nakahara:2003nw} for more details)
\begin{equation}
\frac{dg_\mathcal{i}(t)}{dt}=-\omega(\sigma_{\mathcal{i}*}X)g_\mathcal{i}(t) = -\mathcal{A}_\mathcal{i}(X)g_\mathcal{i}(t),
\end{equation}
where \eqref{sigma and A} has been used.
This leads to
\begin{equation}\label{der of rhoM}
-\frac{d\rho_M(t)}{dt}\rho_M(t)^{-1}
= \bar{\mathcal{A}}_\mathcal{i}(X)
\equiv \frac{dx^\mu}{dt}\bar{\mathcal{A}}_{\mathcal{i}\mu}
\equiv
\frac{dx^\mu}{dt}
\left(
\begin{array}{cc}
\bar{\Gamma}^{(L)}_{\mathcal{i}\mu} & \bar{\Gamma}^{(T)}_{\mathcal{i}\mu} \\
0 & 0 \\
\end{array}
\right)
\end{equation}
according to \eqref{A in Mobius}, and
\begin{equation}\label{part 1}
-\frac{d\rho_V(t)}{dt}\rho_V(t)^{-1}
= \rho_V\!\left(\Gamma^{(L)}_\mathcal{i}(X)\right)
\equiv \frac{dx^\mu}{dt}\rho_V\!\left(\Gamma^{(L)}_{\mathcal{i}\mu}\right).
\end{equation}
Acting \eqref{der of rhoM} on $\bar{\xi}_\mathcal{i}$ via \eqref{Mobius} yields
\begin{equation}\label{part 2}
-\frac{d\rho_M(t)}{dt}\rho_M(t)^{-1}\bar{\xi}_\mathcal{i}(t)
=\frac{dx^\mu}{dt}\left(\bar{\Gamma}^{(L)}_{\mathcal{i}\mu}\bar{\xi}_\mathcal{i} +\bar{\Gamma}^{(T)}_{\mathcal{i}\mu}\right).
\end{equation}
Substituting \eqref{part 1} and \eqref{part 2} into \eqref{exterior covariant derivative}, we have
\begin{equation}
\nabla_X s_{v}
=
\left[
\sigma_i(0),
\left.\frac{dx^\mu}{dt}
\left(
\bar{D}^{(L)}_\mu\bar{\xi}_\mathcal{i}+\bar{\Gamma}^{(T)}_{\mathcal{i}\mu}, D^{(L)}_{V\mathcal{i}\mu}v
\right)\right|_{t=0}
\right],
\end{equation}
where ${dx^\mu(t)}/{dt}:={dx^\mu(\gamma(t))}/{dt}=X^\mu(\gamma(t))$,
or equivalently
\begin{equation}
\nabla s_{v}=
\left[
\sigma_i,
\left(
\bar{D}^{(L)}\bar{\xi}_\mathcal{i}+\bar{\Gamma}^{(T)}_\mathcal{i}, D^{(L)}_{V\mathcal{i}}v
\right)
\right],
\end{equation}
where $D_V^{(L)}$ and $\bar{D}^{(L)}$ are defined in \eqref{DL'} and \eqref{DL bar}.

Remarkably, the new one-form $\theta$ as defined in \eqref{theta and xi again} arises naturally. In terms of $\theta$, the covariant derivative of the section $s_{v}(p)\in\Gamma(U,E)$ can be recast as
\begin{equation}\label{main result 1}
\nabla s_{v}=
\left[
\sigma_\mathcal{i},
\left(\bar{\theta}_\mathcal{i}, D^{(L)}_{V\mathcal{i}}v
\right)
\right],
\end{equation}
or equivalently
\begin{equation}\label{main result 1'}
\nabla_X s_{v}=
\left[
\sigma_\mathcal{i},
\left(\bar{\theta}_\mathcal{i}(X), D^{(L)}_{V\mathcal{i}}(X)v
\right)
\right]
\equiv
\left[
\sigma_\mathcal{i},
\left(\bar{\theta}_\mathcal{i}(X), dv(X)+ \rho_V\big(\Gamma^{(L)}_{\mathcal{i}}(X)\big)v
\right)
\right].
\end{equation}
The one-forms appearing in a local expression for the covariant derivative are to be identified as the local gauge potentials. Therefore, the corresponding local gauge potentials are given by $\theta$ for the ``affine'' part and $\Gamma^{(L)}$ (or, more precisely, the $\rho_V$-representation thereof) for the ``vector'' part.

Since the affine vector $\bar{\xi}$ plays the role of a \emph{reference point} of a local Lorentz frame as discussed in \secref{sec:Poincare symmetry}, the corresponding potential $\theta$ dictates how a reference point is different from its neighbored value via parallel transport. The weak equivalence principle furthermore requires that the parallel transport of a reference point be universal regardless of the matter content. In other words, $\theta$ shall be independent of the vector space $V$, which is chosen to represent the matter field under consideration. Since $\bar{\theta}$ transforms as a vector-valued one-form under a local $A(n,\mathbb{R})$ transformation as indicated in \eqref{theta as a vector} and is independent of the matter content, we must identify $\bar{\theta}$ as the \emph{coframe} one-form (up to an arbitary universal factor).

On the other hand, since $\Gamma^{(L)}$ is insensitive of the translational part of a local $A(n,\mathbb{R})$ transformation as shown in \eqref{gauge transform 1a} or \eqref{gauge transform 2a}, it is identified as the familiar gauge potential associated with $GL(n,\mathbb{R})$.

Having rigorously derived $\theta$ and $\Gamma^{(L)}$ as the local gauge potentials with regard to the covariant derivative of a section on the associated affine-vector bundle, we will study the corresponding curvatures (i.e., field strengths) of the gauge potentials in the next section.

\section{Curvature on the affine-vector bundle}\label{sec:curvature}
In the previous section, we define the covariant derivative $\nabla_X s_v$ with respect to a tangent vector $X\in T_pM$. This leads us to define the corresponding curvature.
To begin with, we compute $\nabla_Y\nabla_X s_{v}$ for any two vector fields $X,Y\in\mathscr{X}(U)$.

Let $\chi:[0,1]\rightarrow U$ be a curve on $U$ whose tangent vectors are given by $Y$, i.e., ${dx^\mu(t)}/{dt}:={dx^\mu(\chi(t))}/{dt}=Y^\mu(\chi(t))$, and $\tilde{\chi}(t)=\sigma_\mathcal{i}(t)g_\mathcal{i}(t)$ be a horizontal lift of $\chi(t)$. Starting from \eqref{main result 1'} and following the same procedures used in \eqref{exterior covariant derivative}, we have
\begin{eqnarray}\label{double derivative}
&&\nabla_Y\nabla_X s_{v}
=\left[\tilde{\chi}(0),
\left.\frac{d}{dt}\left(\rho_A(g_\mathcal{i}(t))^{-1}\left(\bar{\theta}_\mathcal{i}(X(t)), \;\rho_V\big(\Gamma^{(L)}_\mathcal{i}(X(t))\big)v(t)+dv(X(t))\right)\right)\right|_{t=0}
\right] \nonumber\\
&=&\left[\sigma_\mathcal{i}(0),
\left.
\left(
\frac{dx^\mu}{dt}
\rho_{\bar{\theta}}\left(\mathcal{A}_\mathcal{i}(t)\right)\bar{\theta}_\mathcal{i}(X(t))
+\frac{d\bar{\theta}_{\mathcal{i}\mu}(t)}{dt}X^\mu(t)
+\bar{\theta}_{\mathcal{i}\mu}(t)\frac{\partial X^\mu}{\partial x^\nu}\frac{dx^\nu}{dt},
\right. \right. \right.\nonumber\\
&& \qquad \qquad
\left.\left.
\frac{dx^\mu}{dt}\rho_V\big(\Gamma^{(L)}_{\mathcal{i}\mu}(t)\big)\rho_V
\big(\Gamma^{(L)}_\mathcal{i}(X(t))\big)v(t)
+\left(\frac{d}{dt}\rho_V\big(\Gamma^{(L)}_{\mathcal{i}\mu}(t)\big)\right)X^\mu(t)\,v(t)\right.\right. \nonumber\\
&& \qquad \qquad \left.\left.\left. \mbox{}+\rho_V\!\left(\Gamma^{(L)}_{\mathcal{i}\mu}(t)\frac{\partial X^\mu}{\partial x^\nu}\frac{dx^\nu}{dt}\right)v(t) + \frac{d(\partial_\mu v(t))}{dt} X^\mu(t)
+ \partial_\mu v(t)\frac{\partial X^\mu}{\partial x^\nu}\frac{dx^\nu}{dt} \right)\right|_{t=0}\right]\nonumber \\
&=&
\left[
\sigma_\mathcal{i}(0),
\left(\bar{\Gamma}_\mathcal{i}^{(L)}(Y)\bar{\theta}_\mathcal{i}(X) + \frac{1}{2}(Y\!\iprod d\bar{\theta}_\mathcal{i})(X) + \bar{\theta}_{\mathcal{i}\mu} Y[X^\mu],
\right.\right. \nonumber\\
&& \qquad\qquad
\rho_V\!\left(\Gamma^{(L)}_\mathcal{i}(Y)\Gamma^{(L)}_\mathcal{i}(X) + \frac{1}{2} \left(Y\!\iprod d\Gamma_\mathcal{i}^{(L)}\right)(X) +\Gamma_{\mathcal{i}\mu}^{(L)}Y[X^\mu]\right)v \nonumber\\
&&\qquad \qquad \left.\mbox{}+ (\partial_\nu\partial_\mu v)Y^\nu X^\mu + (dv)_\mu Y[X^\mu]
\bigg)\right|_{t=0}
\bigg],
\end{eqnarray}
where ${dx^\mu(t)}/{dt}:={dx^\mu(\chi(t))}/{dt}=Y^\mu(\chi(t))$, $X(t):=X(\chi(t))$, $Y(t):=Y(\chi(t))$, $\mathcal{A}(t):=\mathcal{A}(\chi(t))$, $\bar{\theta}(t):=\bar{\theta}(\chi(t))$, $\Gamma^{(L)}(t):=\Gamma^{(L)}(\chi(t))$, and $\rho_{\bar{\theta}}(\mathcal{A})$ is the representation of $\mathcal{A}$ acting on $\bar{\theta}$, which takes the form
\begin{equation}
\rho_{\bar{\theta}}(\mathcal{A})\, \bar{\theta} = \bar{\Gamma}^{(L)} \bar{\theta}
\end{equation}
in accordance with \eqref{theta as a vector}.

Meanwhile, applying \eqref{main result 1'} again, we have
\begin{equation}\label{lemma 1}
\nabla_{[X,Y]} s_{v}=
\left[
\sigma_\mathcal{i},
\left(\bar{\theta}_\mathcal{i}([X,Y]), D^{(L)}_{V\mathcal{i}}([X,Y])v
\right)
\right].
\end{equation}
Note that $[X,Y]^\mu=X^\nu\partial_\nu Y^\mu-Y^\nu\partial_\nu X^\mu$ implies
\begin{equation}\label{lemma 2}
\eta([X,Y])\equiv \eta_\mu [X,Y]^\mu
=\eta_\mu X[Y^\mu] - \eta_\mu Y[X^\mu]
\end{equation}
for any one-form $\eta$.
Putting \eqref{double derivative}, \eqref{lemma 1}, and \eqref{lemma 2} together, we have
\begin{eqnarray}\label{main result 2}
\mathcal{R}^{(E)}(X,Y)s_{v}
&:=&[\nabla_X,\nabla_Y] s_{v} - \nabla_{[X,Y]} s_{v} \nonumber\\
&=& \left[
\sigma_\mathcal{i},
\left(
\left(d\bar{\theta}_\mathcal{i} + \bar{\Gamma}_i^{(L)}\wedge\bar{\theta}_\mathcal{i}\right)(X,Y),
\rho_V\!\left(\left(d\Gamma_\mathcal{i}^{(L)} +\Gamma_\mathcal{i}^{(L)}\wedge\Gamma_\mathcal{i}^{(L)}\right)(X,Y)\right)v
\right)
\right] \nonumber\\
&=&
\left[
\sigma_\mathcal{i},\left(\bar{T}_\mathcal{i}(X,Y),\rho_V\!\left(R_\mathcal{i}(X,Y)\right)v\right)
\right],
\end{eqnarray}
where $T\in \mathfrak{t}(n,\mathbb{R})\otimes\Omega^2(M)$ is defined as
\begin{equation}\label{def of T}
T := D^{(L)}\theta \equiv d\theta + [\Gamma^{(L)},\theta],
\end{equation}
or equivalently
\begin{equation}\label{def of T bar}
\bar{T} \equiv \rho_n(T):= \bar{D}^{(L)}\bar{\theta} \equiv d\bar{\theta} + \bar{\Gamma}^{(L)}\wedge\bar{\theta},
\end{equation}
and $R\in \mathfrak{gl}(n,\mathbb{R})\otimes\Omega^2(M)$ is defined as
\begin{equation}\label{def of R}
R := d\Gamma^{(L)} + \frac{1}{2}[\Gamma^{(L)},\Gamma^{(L)}] \equiv d\Gamma^{(L)} + \Gamma^{(L)}\wedge\Gamma^{(L)}.
\end{equation}
Note that \eqref{def of T} is identical to the Cartan torsion two-form as defined in \eqref{T and R part a} and \eqref{def of R} identical to the Riemann curvature two-form as defined in \eqref{T and R part b}.

If $X,Y,v$ are replaced by $X'(p)=f(p)X(p)$, $Y'(p)=g(p)Y(p)$, and $v'(p)=h(p)v(p)$, where $f,g,h\in\Omega^0(M)$ are arbitrary scalar fields, it follows from \eqref{main result 2} that
\begin{eqnarray}
\mathcal{R}^{(E)}(X',Y')s_{v'}
&=&
\left[
\sigma_\mathcal{i},\left(\bar{T}_\mathcal{i}(X',Y'), \rho_V\!\left(R_\mathcal{i}(X',Y')\right)v'\right)
\right] \nonumber\\
&=&
\left[
\sigma_\mathcal{i},\left(fg\,\bar{T}_\mathcal{i}(X,Y), fgh\,\rho_V\!\left(R_\mathcal{i}(X,Y)\right)v\right)
\right] \nonumber\\
&\equiv& fg\, \mathcal{R}^{(E)}(X,Y)s_{hv},
\end{eqnarray}
since $T$ and $R$ are two-forms, which by definition are linear when acting on tangent vectors. $\mathcal{R}^{(E)}(X,Y)s_{v}$ is said to be linear in $X$, $Y$, and $v$ in the above sense. Although we define $\mathcal{R}^{(E)}(X,Y)s_{v}$ by considering two vector \emph{fields} $X,Y\in\mathscr{X}(U)$ and a $V$-value \emph{field} $v\in V\otimes\Omega^0(U)$, it turns out that, because of the linearity, $\mathcal{R}^{(E)}(X,Y)s_{v}$ is well defined for any two tangent vectors $X,Y\in T_pM$ and any $V$-value $v\in V$.\footnote{It would not be the case if we define $\mathcal{R}^{(E)}(X,Y)s_{v}$ as $[\nabla_X,\nabla_Y] s_{v}$ without taking into account the term $-\nabla_{[X,Y]} s_{v}$. See Ch.~11 and especially Exercise 11.2 of \cite{Misner:1974qy} for more discussions related to this point.}

\begin{figure}

\centering
 \includegraphics[width=0.35\textwidth]{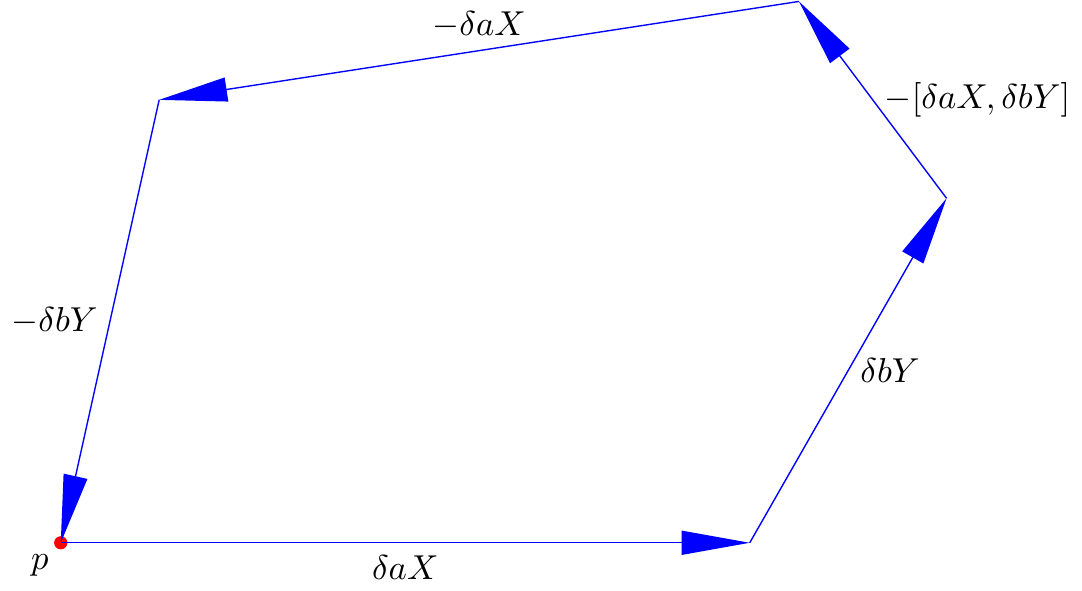}

\caption{A closed curve used to understand the geometric meaning of $\mathcal{R}^{(E)}(\delta aX,\delta bY)s_v$. Given by the two vector fields $X,Y\in\mathscr{X}(U)$, the curve is edged with the vectors $\pm\delta aX$, $\pm\delta bY$, and the negative of the ``closer of the quadrilateral'' $[\delta aX, \delta bY]=\delta a\delta b\,[X, Y]$.\label{fig:holonomy loop}}
\end{figure}

Following the same reasoning explained in Ch.~11 of \cite{Misner:1974qy}, the geometric meaning of the ``curvature'' operator $\mathcal{R}^{(E)}(X,Y)$ acting on a section $s_{v}$ can be understood in terms of holonomy as follows. Consider an infinitesimal closed curve spanned by $\delta aX$ and $\delta bY$ as depicted in \figref{fig:holonomy loop}. Starting at the initial point $p$ with an initial value of $s_v(p)$ as given by \eqref{s v}, if we parallel transport $s_{v}(p)$ around the closed curve until we come back to the point $p$, we end up with a new value $s'_{v'}(p)$. The difference between $s'_{v'}(p)$ and $s_{v}(p)$ is give by
\begin{eqnarray}
s'_{v'}(p) - s_{v}(p)
&\equiv& \left[\sigma_\mathcal{i}(p),(\bar{\xi}'_\mathcal{i}(p)-\bar{\xi}_\mathcal{i}(p),v'-v)\right] \nonumber\\
&=& \mathcal{R}^{(E)}(\delta aX,\delta bY) s_{v} + O(\delta^2) \nonumber\\
&=&\delta a\delta b \left[\sigma_\mathcal{i}, (T_\mathcal{i}(X,Y), \rho_V\big(R_\mathcal{i}(X,Y)\big)v)\right] + O(\delta^2),
\end{eqnarray}
where, as we are comparing $s'_{v'}$ and $s_{v}$ at the same point $p\in U$, we choose the same local section $\sigma_\mathcal{i}\in\Gamma(U,P)$ to represent both $s'_{v'}$ and $s_{v}$.

Under the parallel transport around an infinitesimal closed curve, $R$ gives a value of $\mathfrak{gl}(n,\mathbb{R})$ that dictates how a given vector $v\in V$ is linearly changed to $v'$. The geometric picture of $R$ in terms of holonomy in the ``vector'' part of the affine-vector bundle is exactly the same as that of the standard $GL(n,\mathbb{R})$ Yang-Mills theory on a conventional vector bundle.
Analogously, on the other hand, $T$ gives a value of $\mathfrak{t}(n,\mathbb{R})$ that dictates how a given reference point $\bar{\xi}$ is displaced to $\bar{\xi}'$ under the parallel transport around the closed curve. This provides a clear geometric picture of $T$ in terms of holonomy in the ``affine'' part of the affine-vector bundle.

Also note that whereas the local section $s_v\in\Gamma(U_\mathcal{i},E)$ in \eqref{s v} involves two gauge choices --- $\sigma_\mathcal{i}\in\Gamma(U_\mathcal{i},P)$ and $\bar{\xi}_\mathcal{i}:U_\mathcal{i}\rightarrow\mathbb{R}^n$, the honolomy of $s_v$ around an infinitesimal closed curve is given by $T$ and $R$, both of which are \emph{covariant} under a local $GL(n,\mathbb{R})$ transformation and \emph{invariant} under a local translation. (Notably, the honolomy of $s_v$ around a closed curve is completely independent of the gauge ambiguity of $\bar{\xi}$.) This suggests that both the holonomies in the ``affine'' and ``vector'' parts are physical and yield observational consequences, which will be discussed in the next section.

\section{Observational consequences}\label{sec:observational consequences}
The observational consequences of the Reimann curvature $R$ are well understood. Typically, $R\neq 0$ imposes the geodetic effect and frame-dragging on the dynamics of a matter field (see \cite{Everitt:2011hp} for the experimental verification). Additionally, even if the trajectories of a matter field are confined to a region where $R=0$, a two-path interference experiment will still yield different interference patterns \textit{\`{a} la} the Aharonov-Bohm effect, in response to the holonomy of $\Gamma^{(L)}$ along the closed curve composed of the two worldline paths. The holonomy of $\Gamma^{(L)}$ along a finite closed curve $\gamma$ is defined as the pathwise integral $\mathcal{P}\,e^{\oint_\gamma \Gamma^{(L)}}$, which is equal to $e^{\oiint_\gamma R}$ --- i.e.\ the exponential of the flux integral of $R$ over the surface enclosed by $\gamma$ --- according to \eqref{main result 2}. It is possible to have $R=0$ everywhere along $\gamma$ but still have nontrivial holonomy, thus yielding an observable result of a nontrivial interference pattern \cite{dowker1967gravitational, bezerra1987gravitational, bezerra1990some}.

On the other hand, the observational consequences of the Cartan torsion $T$ are much less understood.
In the theories of first-order Lagrangians, it has been shown that torsion is inextricably bound to spinning matter and vanishes wherever there is no spinning matter \cite{hehl1976general}. Consequently, it does not propagate through vacuum and only gives rise to a ``direct interaction'' in contact with spin source for the dynamics of a matter field.
Even in the theories of second-order Lagrangians including torsion-torsion coupling, the dynamics of torsion does not change in an essential way (see \cite{von1975equivalence} for more discussions). Therefore, the Cartan torsion has no ``at-a-distance'' effects analogous to the geodetic effect and frame-dragging of the Riemann curvature.

Nevertheless, the geometric meaning of $T$ in terms of holonomy leads us to anticipate a kinematical effect of torsion that in principle can be measured \textit{\`{a} la} the Aharonov-Bohm effect, analogous to the case of $R$.
If the surface enclosed by the two paths contains spinning matter, we may have nontrivial holonomy $\mathcal{P}\,e^{\oint_\gamma \theta} \equiv e^{\oint_\gamma \theta}=e^{\oiint_\gamma T}$ according to \eqref{main result 2}.\footnote{Because $[P_i,P_j]=0$, the pathwise symbol $\mathcal{P}$ here is superfluous and consequently we have $\oint_\gamma \theta=\oiint_\gamma T$.} The logarithm of the holonomy, $\oint_\gamma \bar{\theta}=\oiint_\gamma \bar{T}$, is to be understood as the displacement vector between the initial reference point and the final reference point under the parallel transport around $\gamma$. Although the gauge choice of the reference point $\bar{\xi}$ is arbitrary, the displacement vector associated with the holonomy is non-arbitrary and physical. Consider that a local Lorentz frame starting at a given point is parallel transported along two worldline paths separately. If the two paths join again at a destination point, the two transported Lorentz frames overlap but their coordinate origins might not coincide, and the difference between the frame origins is described by the displacement vector of the holonomy along the closed curve composed of the two paths. Since the interference experiment measures the two wavefunctions along the two paths superimposed at the destination point, the difference between the origins of the two overlapped frames and consequently the displacement vector of the honolomy should be manifested in the interference pattern.
Therefore, despite the fact that torsion only gives a direct-contact interaction, its flux can still be measured \textit{\`{a} la} the Aharonov-Bohm effect without being probed directly.

It should be emphasized that the conjecture given above is based purely on \emph{kinematical} considerations and remains speculative. To ensure that this effect is indeed physically measurable, one has to take into account \emph{dynamical} considerations to see whether the dynamics dictates that the Lorentz frames are parallel transported in accordance with the $A(n,\mathbb{R})$-connection. More precisely, the Lorentz frames should be operationally defined with reference to some physical objects, namely, matter fields, and in principle it is the dynamics of matter fields that dictates how the Lorentz frames are transported.

For the dynamics of matter fields in the gauge theory of a general $G$-connection (with the gauge group $G$ being different from $SO(1,3)$), it has been shown that the equations of motion of ordinary matter fields do not follow the parallel transport law in accordance with the $G$-connection, but instead they follow the parallel transport law either of the Levi-Civita connection (i.e., the Riemannian connection of the metric) alone or of a hybrid connection resulting from both the Levi-Civita connection and the general $G$-connection \cite{Audretsch:1981xn,Cembranos:2018ipn,Hayashi:1990ig,Bergmann:1980wt,hehl1973spin}.
Furthermore, it has been argued that, by starting from the action principle, there is no clear way to formulate the dynamics of matter fields whereby the matter fields follow the parallel transport law of the $G$-connection  (see Section VI.C of \cite{Jim_nez_2020}).

The ``no-go'' results from dynamical considerations of matter fields seem to suggest that the conjectured Aharonov-Bohm-like effect is not measurable after all. However, we should bear in mind that, for the matter dynamics studied so far in the literature, a matter field $\varphi(x^\mu)$ is essentially treated as a local section of an associated vector bundle $P\times_\rho V$, where $V$ is the vector space in which $\varphi(x^\mu)$ resides and $\rho$ is the linear presentation of $G$ over $V$ (recall \footref{foot:field as a local section} for the case of $G=SO(1,n-1)$). This treatment does not faithfully reflect the affine structure of $A(n,\mathbb{R})$ as argued in \secref{sec:Poincare symmetry}, and cannot make sense of an affine vector $\bar{\xi}$ as a reference point (recall \footref{foot:vector bundle}).
The results of our work suggest that a more suitable arena for the gauge theory of $A(n,\mathbb{R})$ is given by an associated affine-vector bundle $P\times_{\rho_A}F$, where $F=\mathbb{R}^n\times V$ is the affine-vector space and $\rho_A$ is the affine (nonlinear) representation of $A(n,\mathbb{R})$. If we attempt to formulate the dynamical theory of matter fields in the framework of an affine-vector bundle, a matter field should be treated as a local section of $P\times_{\rho_A}F$, instead of that of $P\times_\rho V$. That is, a matter field is locally described by the pair $(\bar{\xi}(x^\mu),\varphi(x^\mu))$, rather than $\varphi(x^\mu)$ alone. With the inclusion of the accompanying field $\bar{\xi}(x^\mu)$, the resulting dynamics of matter fields could be qualitatively different from the results that have been studied so far. Therefore, it would be premature to jump to the conclusion that the Aharonov-Bohm-like effect is not physically measurable.

Of course, it remains a challenging open question how the dynamical theory of matter fields should be formulated through affine-vector bundles. Especially, it is unclear how the reference variable $\bar{\xi}(x^\mu)$ should be included in the action. Should it be treated as a dynamical variable or merely an algebraic one? Depending on the details of formulation, the dynamics might result in a complicated form of transport upon the Lorentz frames. However complicated, nevertheless, it is still likely to give rise to a measurable Aharonov-Bohm-like effect, although the detailed form of the effect could be quantitatively different from what we conjectured solely from kinematical considerations.
As regards the dynamics of matter fields, one important issue in relation to the weak equivalence principle is whether the variable $\bar{\xi}(x^\mu)$ should be treated as universal for different matter species or different matter species should be associated with different $\bar{\xi}(x^\mu)$ variables. If the former is the case, the displacement vector inferred from the interference pattern of the Aharonov-Bohm-like effect is expected to be identical regardless of the matter species used for the experiment. If the latter is the case, the weak equivalence principle is not fully satisfied in the sense that we cannot directly identify $\bar{\theta}$ appearing in \eqref{main result 1} as the coframe one-form, or alternatively different matter species are said to see intricately different spacetime structures of different coframe one-forms.
In any case, studying the dynamics in the framework of an affine-vector bundle will shed new light on and open new possibilities for various important issues that have been studied in the literature.

\section{Summary and discussion}\label{sec:summary}
Investigating the Poincar\'{e} symmetry more carefully in view of the Einstein equivalence principle motivates us to propose the framework of an associated affine-vector bundle, which contains an affine vector $\bar{\xi}$ on the fiber serving as a reference point and thus provides a more suitable arena for the affine group $A(n,\mathbb{R})$. The associated affine-vector bundle is rigorously defined in the same spirit of the formal quotient-space construction for an associated vector bundle.
Choosing a local section $s_v\in\Gamma(U,E)$ on the affine-vector bundle $E$ amounts to the familiar gauge fixing of choosing a local section $\sigma_\mathcal{i}\in\Gamma(U,P)$ on the principle bundle $P$ plus the extra gauge fixing of choosing an arbitrary reference point $\bar{\xi}_\mathcal{i}$,

The formal definitions of the parallel transport and the covariant derivative on an associated vector bundle in the Ehresmann-connection approach can be naturally generalized to the associated affine-vector bundle. Rigorously deriving the covariant derivative of a local section $s_v$ on the affine-vector bundle, we obtain the result in \eqref{main result 1} and \eqref{main result 1'}. Remarkably, $\theta$ and $\Gamma^{(L)}$ naturally appear as the gauge potentials for the ``affine'' part and the ``vector'' part, respectively. Because $\bar{\theta}$ transforms as a vector under a local $A(n,\mathbb{R})$ transformation and the weak equivalence principle suggests that $\bar{\theta}$, being a measure of the parallel transport of a reference point, has to be universal regardless of the matter content, the gauge potential $\theta$ is to be identified as the coframe one-form field.

Applying the covariant derivative twice on the local section $s_v$, we rigorously derive the curvature operator $\mathcal{R}^{(E)}(X,Y)$ acting on $s_v$ as given in \eqref{main result 2}. As expected, the Cartan torsion two-form $T$ and the Riemann curvature two-form $R$ naturally appear as the gauge field strengths for the ``affine'' part and the ``vector'' part, respectively. The geometric meanings of $T$ and $R$ in terms of holonomy become clear.

Our approach rigorously derives the appealing parallel between $R$ and $T$ from first principles without any ad hoc prescriptions, and provides a clear geometric and physical picture of them. Believing that the affine-vector bundle is more fundamental than a conventional vector bundle on the grounds that any local Lorentz frame shall manifest the full symmetry of the Poincar\'{e} group including its translational part, we arrive at a conjecture about a kinematical effect of the Cartan torsion that in principle can be measured \textit{\`{a} la} the Aharonov-Bohm effect. The exact quantitative prediction of this effect, however, might depend on the dynamical theory of matter fields.

It should be remarked that, as noted in the last sentence of \secref{sec:Poincare symmetry}, our analysis is carried out in the broader context of the affine group $A(n,\mathbb{R})$, but it has not addressed the issue of how the gauge group $A(n,\mathbb{R})$ is reduced to the Poincar\'{e} group. In other words, we view the coframe field $\bar{\theta}$ defined in \eqref{coframe} simply as a universal vector-valued one-form, but so far have not takeen into account the important fact that $\theta$ gives rise to the spacetime metric tensor via
\begin{equation}
g_{\mu\nu} = \eta_{ij}{e^i}_\mu{e^j}_\nu.
\end{equation}
It is curious why the coframe field has two very different geometric meanings: as the gauge potential associated with translation and as the ``square root'' of the metric tensor. Some theories that include the \emph{nonmetricity} field as an additional dynamical variable  \cite{blagojevic2002gravitation, mielke2017geometrodynamics, gronwald1996gauge} might eventually provide a dynamical explanation for the dual role of $\theta$.
Although this paper does not consider this question at all, the clear geometric and physical picture we have obtained in the framework of an associated affine-vector bundle might offer valuable new insight about the intriguing relation between the local translational symmetry and the spacetime metric.

Finally, we comment that the attention of this paper is focused on the kinematical aspects of MAG, and we leave the dynamical aspects of formulating the Lagrangian of MAG for future research.
Akin to what we have observed for the covariant Dirac equation and the covariant Dirac Lagrangian in \secref{sec:Poincare symmetry}, various Lagrangian actions of gravity, particularly the Einstein-Cartan action, formulated in the context of MAG are invariant under local Lorentz transformations as well as under diffeomorphisms but \emph{not under local translations} (see Sec.~3.2.3 of \cite{blagojevic2002gravitation}). In this sense, they are not really cast as ``true'' gauge theories of the affine group. According to Trautman's idea that the affine-vector-valued field $\bar{\xi}$ can be viewed as a ``generalized Higgs field'' \cite{Trautman:1979cq}, perhaps it is possible to formulate an explicit spontaneous symmetry breaking mechanism responsible for ``hiding'' the local translational symmetry in the Lagrangian action. There have been various approaches in this direction (see e.g.\ \cite{Wise:2006sm} and the recent work \cite{Obukhov:2020uan}), but a satisfactory theory is still missing. The formalism of the associated affine-vector bundle we have devised might provide a better mathematical framework for implementing such a theory.
As has been remarked in the end of \secref{sec:observational consequences}, investigation of the dynamical theories in the framework of an affine-vector bundle will offer new insight into various important issues in the literature.


\begin{acknowledgments}
The authors would like to thank two anonymous reviewers for their valuable suggestions, which have helped to improve the manuscript greatly.
This work was supported in part by the Ministry of Science and Technology, Taiwan under the Grants MOST 107-2119-M-002-031-MY3 and MOST 109-2112-M-110-021.
\end{acknowledgments}







\end{document}